\documentclass[sigconf]{acmart}

\usepackage{booktabs} 

\usepackage{url}
\usepackage{epsfig}
\usepackage{bm}
\usepackage{balance}
\usepackage{acronym}
\usepackage{siunitx}
\usepackage{subcaption}
\usepackage{colortbl}
\usepackage{enumitem}

\copyrightyear{2020} 
\acmYear{2020} 
\setcopyright{acmcopyright}\acmConference[SIGIR '20]{Proceedings of the 43rd International ACM SIGIR Conference on Research and Development in Information Retrieval}{July 25--30, 2020}{Virtual Event, China}
\acmBooktitle{Proceedings of the 43rd International ACM SIGIR Conference on Research and Development in Information Retrieval (SIGIR '20), July 25--30, 2020, Virtual Event, China}
\acmPrice{15.00}
\acmDOI{10.1145/3397271.3401036}
\acmISBN{978-1-4503-8016-4/20/07}

\acrodef{ER}[ER]{Effect Ratio}
\acrodef{DeltaRI}[DeltaRI]{Delta Relative Improvement}

\definecolor{mm}{RGB}{127,201,127}
\definecolor{nf}{RGB}{190,174,212}
\definecolor{tb}{RGB}{253,192,134}
\definecolor{ps}{RGB}{153,50,204}
\definecolor{td}{RGB}{56,108,176}

\definecolor{pt}{RGB}{175,238,238}
\definecolor{lg}{RGB}{211,211,211}

\begin{document}

\title{How to Measure the Reproducibility of \\ System-oriented IR Experiments}

\author{Timo Breuer}
\affiliation{%
\institution{TH K{\"o}ln, Germany}
}
\email{timo.breuer@th-koeln.de}

\author{Nicola Ferro}
\affiliation{%
\institution{University of Padua, Italy}
}
\email{ferro@dei.unipd.it}

\author{Norbert Fuhr}
\affiliation{%
\institution{Universit{\"a}t Duisburg-Essen, Germany}
}
\email{norbert.fuhr@uni-due.de}

\author{Maria Maistro}
\affiliation{%
\institution{University of Copenhagen, Denmark}
}
\email{mm@di.ku.dk}

\author{Tetsuya Sakai}
\affiliation{%
\institution{Waseda University, Japan}
}
\email{tetsuyasakai@acm.org}

\author{Philipp Schaer}
\affiliation{%
\institution{TH K{\"o}ln, Germany}
}
\email{philipp.schaer@th-koeln.de}

\author{Ian Soboroff}
\affiliation{%
\institution{NIST, USA}
}
\email{ian.soboroff@nist.gov}


\renewcommand{\shortauthors}{T. Breuer et. al.}

\renewcommand{\acffont}[1]{\textsl{#1}}

\begin{abstract}
Replicability and reproducibility of experimental results are primary concerns in all the areas of science and \acs{IR} is not an exception. Besides the problem of moving the field towards more reproducible experimental practices and protocols, we also face a severe methodological issue: we do not have any means to assess \emph{when reproduced is reproduced}. Moreover, we lack any reproducibility-oriented dataset, which would allow us to develop such methods.

To address these issues, we compare several measures to objectively quantify to what extent we have replicated or reproduced a system-oriented IR experiment. These measures operate at different levels of granularity, from the fine-grained comparison of ranked lists, to the more general comparison of the obtained effects and significant differences. Moreover, we also develop a reproducibility-oriented dataset, which allows us to validate our measures and which can also be used to develop future measures.
\end{abstract}



\maketitle


\section{Introduction}
\label{sec:introduction}

We are today facing the so-called \emph{reproducibility crisis}~\cite{Baker2016,OSC2015} across all areas of science, where researchers fail to reproduce and confirm previous experimental findings. This crisis obviously involves also the more recent computational and data-intensive sciences~\cite{zz-DagstuhlSeminar16041,NAP2019}, including hot areas such as artificial intelligence and machine learning~\cite{Gibney2020}. For example, \citet{Baker2016} reports that roughly 70\% of researchers in physics and engineering fail to reproduce someone else's experiments and roughly 50\% fail to reproduce even their own experiments.

\ac{IR} is not an exception and researchers are paying more and more attention to what the reproducibility crisis may mean for the field, even more with the raise of the new deep learning and neural approaches \cite{Crane2018,DacremaEtAl2019}.

In addition to all the well-known barriers to reproducibility~\cite{zz-DagstuhlSeminar16041}, a fundamental methodological question remains open: \emph{When we say that an experiment is reproduced, what exactly do we mean by it?} The current attitude is some sort of ``\emph{close enough}'': researchers put any reasonable effort to understand how an approach was implemented and how an experiment was conducted and, after some (several) iterations, when they obtain performance scores which somehow resemble the original ones, they decide that an experimental result is reproduced. Unfortunately, \ac{IR} completely lacks any means to \emph{objectively measure} when reproduced is reproduced and this severely hampers the possibility both to assess to what extent an experimental result has been reproduced and to sensibly compare among different alternatives for reproducing an experiment. 

This severe methodological impediment is not limited to \ac{IR} but it has been recently brought up as a research challenge also in the 2019 report on ``Reproducibility and Replicability in Science'' by the US \citet[p.~62]{NAP2019}: ``The National Science Foundation should consider investing in research that explores the limits of computational reproducibility in instances in which bitwise reproducibility\footnote{``For computations, one may expect that the two results be identical (i.e., obtaining a bitwise identical numeric result). In most cases, this is a reasonable expectation, and the assessment of reproducibility is straightforward. However, there are legitimate reasons for reproduced results to differ while still being considered consistent''~\cite[p.~59]{NAP2019}. The latter is clearly the most common case in \ac{IR}.} is not reasonable in order to ensure that the \emph{meaning of consistent computational results} remains in step with the development of new computational hardware, tools, and methods''.
Another severe issue is that we lack any \emph{experimental collection} specifically focused on reproducibility and this prevents us from developing and comparing measures to assess the extent of achieved reproducibility.

In this paper, we tackle both these issues. Firstly, we consider different measures which allow for comparing experimental results at different levels from most specific to most general: the ranked lists of retrieved documents; the actual scores of effectiveness measures; the observed effects and significant differences.
As you can note these measures progressively depart more and more from the ``bitwise reproducibility''~\cite{NAP2019} which in the \ac{IR} case would mean producing exactly identical ranked lists of retrieved documents. 
Secondly, starting from TREC data, we develop a reproducibility-oriented dataset and we use it to compare the presented measures. 

The paper is organized as follows: Section~\ref{sec:related} discusses related work; Section~\ref{sec:framework} introduces the evaluation measures under investigation; Section~\ref{sec:dataset} describes how we created the reproducibility-oriented dataset; Section~\ref{sec:experiments_new} presents the experimental comparison of the evaluation measures; finally, Section~\ref{sec:conclusions} draws some conclusions and outlooks future work.

\section{Related Work}
\label{sec:related}

In defining what repeatability, replicability, reproducibility, and other of the so-called \emph{r-words} are~\cite{Plesser2018}, \citet{DeRoure2014} lists 21 r-words grouped in 6 categories, which range from scientific method to understanding and curation. In this paper, we broadly align with the definition of replicability and reproducibility currently adopted by the \ac{ACM}\footnote{\url{https://www.acm.org/publications/policies/artifact-review-badging}, April 2018.}:
\begin{itemize}[leftmargin=*]
\item \emph{Replicability (different team, same experimental setup)}: the measurement can be obtained with stated precision by a different team using the same measurement procedure, the same measuring system, under the same operating conditions, in the same or a different location on multiple trials. For computational experiments, an independent group can obtain the same result using the author’s own artifacts;
\item \emph{Reproducibility (different team, different experimental setup)}: the measurement can be obtained with stated precision by a different team, a different measuring system, in a different location on multiple trials. For computational experiments, an independent group can obtain the same result using artifacts which they develop completely independently.
\end{itemize}

\subsection*{Reproducibility Efforts in IR}

There have been and there are several initiatives related to reproducibility in \ac{IR}. Since 2015, the ECIR conference hosts a track dedicated to papers which reproduce existing studies, and all the major IR conferences ask an assessment of the ease of reproducibility of a paper in their review forms. The SIGIR group has started a task force~\cite{FerroKelly2018} to define what reproducibility is in system-oriented and user-oriented \ac{IR} and how to implement the \ac{ACM} badging policy in this context. \citet{Fuhr2017,Fuhr2019} urged the community to not forget about reproducibility and discussed reproducibility and validity in the context of the CLEF evaluation campaign. The recent ACM JDIQ special issue on reproducibility in \ac{IR}~\cite{FerroEtAl2018g,FerroEtAl2018h} provides an updated account of the state-of-the-art in reproducibility research as far as evaluation campaigns, collections, tools, infrastructures and analyses are concerned. The SIGIR 2015 RIGOR workshop~\cite{ArguelloEtAl2015} investigated reproducibility, inexplicability, and generalizability of results and held a reproducibility challenge for open source software~\cite{LinEtAl2016}. The SIGIR 2019 OSIRRC workshop~\cite{ClancyEtAl2019} conducted a replicability challenge based on Docker containers.

CENTRE\footnote{\url{https://www.centre-eval.org/}} is an effort  across CLEF~\cite{FerroEtAl2018e,FerroEtAl2019c}, TREC~\cite{SoboroffEtAl2018b}, and NTCIR~\cite{SakaiEtAl2019b} to run a joint evaluation activity on reproducibility. One of the goals of CENTRE was to define measures to quantify to which extent experimental results were reproduced. However, the low participation in CENTRE prevented the development of an actual reproducibility-oriented dataset and hampered the possibility of developing and validating measures for reproducibility. 

\subsection*{Measuring Reproducibility}

To measure reproducibility, CENTRE exploited: Kendall's $\tau$~\cite{Kendall1948}, to measure how close are the original and replicated list of documents; \ac{RMSE}~\cite{KenneyKeeping1954}, to quantify how close are the effectiveness scores of the original and replicated runs; and the \acf{ER}~\cite{SakaiEtAl2019b}, to quantify how close are the effects of the original and replicated/reproduced systems.

We compare against and improve with respect to previous work within CENTRE. Indeed, Kendall's $\tau$ cannot deal with rankings that do not contain the same elements; CENTRE overcomes this issue by considering the union of the original and replicated rankings and comparing with respect to it; we show how this is a somehow pessimistic approach, penalizing the systems and propose to use \acf{RBO}~\cite{WebberEtAl2010}, since it is natively capable to deal with rankings containing different elements. Furthermore, we complement \acf{ER} with the new \acf{DeltaRI} score, to better grasp replicability and reproducibility in terms of absolute scores and to provide a visual interpretation of the effects. Finally, we propose to test replicability and reproducibility with paired and unpaired t-test~\cite{Student1908} respectively, and to use $p$-values as an estimate of replicability and reproducibility success.

To the best of our knowledge, inspired by the somehow unsuccessful experience of CENTRE, we are the first to systematically investigate measures for guiding replicability and reproducibility in \ac{IR}, backing this with the development of a reproducibility-oriented dataset. As previously observed, there is a compelling need for reproducibility measures for computational and data-intensive sciences~\cite{NAP2019}, being the largest body of knowledge focused on traditional lab experiments and metrology~\cite{NAP2016,ISO-5725-2:2019}, and we try here to start addressing that need in the case of \ac{IR}.

\section{Proposed Measures}
\label{sec:framework}

We first introduce our notation. In all cases we assume that the original run $r$ is available. For replicability (\S~\ref{subsec:replicability}), both the original run $r$ and the replicated run $r'$ contain documents from the original collection $C$. For reproducibility (\S~\ref{subsec:reproducibility}), $r$ denotes the original run on the original collection $C$, while $r'$ denotes the reproduced run on the new collection $D$. Topics are denoted by $j \in \{1, \ldots , n_C\}$ in $C$ and $j \in \{1, \ldots , n_D\}$ in $D$, while rank positions are denoted by $i$.
$M$ is any \ac{IR} evaluation measure e.g.,~P@10, \acs{AP}, \acs{nDCG}, where the superscript $C$ or $D$ refers to the collection. $M^{C}(r)$ is the vector of length $n_C$ where each component, $M_j^{C}(r)$, is the score of the run $r$ with respect to the measure $M$ and topic $j$. $\overline{M^{C}(r)}$ is the average score computed across topics.

\subsection{Replicability}
\label{subsec:replicability}

We evaluate replicability at different levels: (i) we consider the actual \emph{ordering of documents} by using Kendall's $\tau$ and \acf{RBO}~\cite{WebberEtAl2010}; (ii) we compare the runs in terms of \emph{effectivenes} with \ac{RMSE}; (iii) we consider whether the \emph{overall effect} can be replicated with \acf{ER} and \acf{DeltaRI}; and (iv) we compute \emph{statistical comparisons} and consider the $p$-value of a paired t-test. While Kendall's $\tau$, \ac{RMSE} and \ac{ER} were originally proposed for CENTRE, the other approaches has never been used for replicability.

It is worth mentioning that these approaches are presented from the most specific to the most general. Kendall's $\tau$ and \ac{RBO} compares the runs at document level, \ac{RMSE} accounts for the performance at topic level, \ac{ER} and \ac{DeltaRI} focus on the overall performance by considering the average across topics, while the $t$-test can just inform us on the significant differences between the original and replicated runs. Moreover, perfect equality for Kendall's $\tau$ and \ac{RBO} implies perfect equality for \ac{RMSE}, \ac{ER}/\ac{DeltaRI} and $t$-test, and perfect equality for \ac{RMSE} implies perfect equality for \ac{ER}/\ac{DeltaRI} and $t$-test, while viceversa is in general not true.

As reference point, we consider the average score across topics of the original and replicated runs, called  \acf{ARP}. Its delta represents the current ``naive'' approach to replicability, simply contrasting the average scores of the original and replicated runs. 




\subsubsection*{Ordering of Documents}
Kendall's $\tau$ is computed as follows~\cite{Kendall1948}:
\begin{equation}
 	\begin{aligned}
 		\tau_j(r, r') & = \frac{P-Q}{\sqrt{\big(P + Q + U\big)\big(P + Q + V\big) }} \\
 		\bar{\tau}(r, r') & = \frac{1}{n_C} \sum_{j = 1}^{n_C} \tau_j (r, r')
 	\end{aligned}
 	\label{eq:tau}
 \end{equation}	
where $\tau_j(r, r')$ is Kendall's $\tau$ for the $j$-th topic, $P$ is the total number of concordant pairs (document pairs that are ranked in the same order in both vectors), $Q$ the total number of discordant pairs (document pairs that are ranked in opposite order in the two vectors), $U$ and $V$ are the number of ties, in $r$ and $r'$ respectively.

This definition of Kendall's $\tau$ is originally proposed for permutations of the same set of items, therefore it is not directly applicable whenever two rankings do not contain the same set of documents. However, this is not the case of real runs, which often return different sets of documents. Therefore, as done in CENTRE@CLEF~\cite{FerroEtAl2018e,FerroEtAl2019c}, we consider the correlation with respect to the union of the rankings. We refer to this method as \emph{Kendall's $\tau$ Union}. The underlying idea is to compare the relative orders of documents in the original and replicated rankings. For each topic, we consider the union of $r$ and $r'$, by removing duplicate entries. Then we consider the rank positions of documents from the union in $r$ and $r'$, obtaining two lists of rank positions. Finally, we compute the correlation between these two lists of rank positions. Note that, whenever two rankings contain the same set of documents, Kendall's $\tau$ in Eq.~\eqref{eq:tau} and Kendall's $\tau$ Union are equivalent. To better understand how Kendall's $\tau$ Union is defined, consider two rankings: $r = [d_1, d_2, d_3]$ and $r' = [d_1, d_2, d_4]$, the union of $r$ and $r'$ is $[d_1, d_2, d_3, d_4]$, then the two lists of rank positions are $[1, 2, 3]$ and $[1, 2, 4]$ and the final Kendall's $\tau$ is equal to $1$. Similarly consider $r = [d_1, d_2, d_3, d_4]$ and $r' = [d_2, d_5, d_3, d_6]$, the union of $r$ and $r'$ is $[d_1, d_2, d_3, d_4, d_5,$ $d_6]$, then the two lists of rank positions are $[1, 2, 3, 4]$ and $[2, 5, 3, 6]$ and the final Kendall's $\tau$ is equal to $2/3$. 




We also consider Kendall's $\tau$ on the intersection of the rankings instead of the union. As reported in~\cite{SandersonSoboroff2007}, Kendall's $\tau$ can be very noisy with small rankings and should be considered together with the size of the overlap between the $2$ rankings. However, this approach does not inform us on the rank positions of the common documents. Therefore, to seamlessly deal with rankings possibly containing different documents and to accout for their rank positions, we propose to use \acf{RBO}~\cite{WebberEtAl2010}, which assumes $r$ and $r'$ to be infinite runs:
\begin{equation}
     \begin{aligned}
     \text{RBO}_j(r,r') &= (1 - \phi) \sum_{i=1}^{\infty}\phi^{i - 1} \cdot A_{i} \\
     \overline{\text{RBO}}(r,r') & = \frac{1}{n_C} \sum_{j = 1}^{n_C} \text{RBO}_j (r, r')
     \end{aligned}
     \label{eq:rbo}
 \end{equation}
where $\text{RBO}_j(r,r')$ is \ac{RBO} for the $j$-th topic; $\phi \in [0, 1]$ is a parameter to adjust the measure top-heaviness: the smaller $\phi$, the more top-weighted the measure; and $A_{i}$ is the proportion of overlap up to rank $i$, which is defined as the cardinality of the intersection between $r$ and $r'$ up to $i$ divided by $i$. 
Therefore, \ac{RBO} accounts for the overlap of two rankings and discounts the overlap while moving towards the end of the ranking, since it is more likely for two rankings to have a greater overlap when many rank positions are considered. 

\subsubsection*{Effectiveness}
As reported in CENTRE@CLEF~\cite{FerroEtAl2018e,FerroEtAl2019c}, we exploit \acf{RMSE}~\cite{KenneyKeeping1954} to measure how close the effectiveness scores of the replicated and original runs are: 
\begin{equation}
	\mathrm{RMSE}\left(M^C(r), M^C(r')\right) = \sqrt {\frac{1}{n_C} \sum_{j = 1}^{n_C} \big(M_{j}^C(r) - M_{j}^C(r') \big)^2}
	\label{eq:rmse}
\end{equation}
\ac{RMSE} depends just on the evaluation measure and on the relevance label of each document, not on the actual documents retrieved by each run. Therefore, if two runs $r$ and $r'$ retrieve different documents, but with the same  relevance labels, then \ac{RMSE} is not affected and returns a perfect replicability score equal to $0$; on the other hand, Kendall's $\tau$ and \ac{RBO} will be able to detect such differences.

Although \ac{RMSE} and the naive comparison of \ac{ARP} scores can be thought as similar approaches, by taking the squares of the absolute differences, \ac{RMSE} penalizes large errors more. This can lead to different results, as shown in Section~\ref{sec:experiments_new}.

\subsubsection*{Overall Effect}
In this case, we define a replication task from a different perspective, as proposed in CENTRE@NTCIR~\cite{SakaiEtAl2019b}.
Given a pair of runs, $a$ and $b$, such that the advanced $a$-run has been reported to outperform the baseline $b$-run on the collection $C$,
can another research group replicate the improvement of the advanced run over the baseline run on $C$?
%
With this perspective, the per-topic improvements in the original and replicated experiments 
are:
\begin{equation}
\Delta M_{j}^{C} = M_{j}^{C}(a) - M_{j}^{C}(b) \ , \,\,\,\,\, \Delta' M_{j}^{C} = M_{j}^{C}(a') - M_{j}^{C}(b')
\label{eq:per-topic-improvements}
\end{equation}
where $a'$ and $b'$ are the replicated advanced and baseline runs respectively.
Note that even if the $a$-run outperforms the $b$-run on average, the opposite may be true for some topics: that is, per-topic improvements may be negative.

Since IR experiments are usually based on comparing mean effectiveness scores, \acf{ER}~\cite{SakaiEtAl2019b} focuses on the replicability of the overall effect as follows:
\begin{equation}
\text{ER}\left(\Delta' M^{C}, \Delta M^{C}\right) = 
\frac{
\overline{\Delta' M^{C}}
}{
\overline{\Delta M^{C}}
}
=
\frac{
\frac{1}{n_{C}}\sum_{j=1}^{n_{C}} \Delta' M_{j}^{C}
}{
\frac{1}{n_{C}}\sum_{j=1}^{n_{C}} \Delta M_{j}^{C}
}
\end{equation}
where the denominator of \ac{ER} is the mean improvement in the original experiment,
while the numerator is the mean improvement in the replicated experiment.
Assuming that the standard deviation for the difference in terms of measure $M$ is common across experiments,
ER is equivalent to the ratio of \emph{effect sizes} (or \emph{standardised mean differences} for the \emph{paired data} case)~\cite{Sakai2018}: 
hence the name.

$\text{ER} \leq 0$ means that the replicated $a$-run failed to outperform the replicated $b$-run: the replication is a complete failure. If $0 < \text{ER} <1$, the replication is somewhat successful, but the effect is smaller compared to the original experiment. If $\text{ER}=1$, the replication is perfect in the sense that the original effect has been recovered as is. If $\text{ER}>1$, the replication is successful, and the effect is actually larger compared to the original experiment.

Note that having the same mean delta scores, i.e.~$ER = 1$, does not imply that the per-topic replication is perfect. For example, consider two topics $i$ and $j$ and assume that the original delta scores are $\Delta M_{i}^{C} = 0.2$ and $\Delta M_{j}^{C} = 0.8$ while the replicated delta scores are $\Delta'M_{i}^{C} = 0.8$ and $\Delta'M_{j}^{C} = 0.2$. Then \ac{ER} for this experiment is equal to $1$. While this difference is captured by \ac{RMSE} or Kendall's $\tau$, which focus on a per-topic level, \ac{ER} considers instead whether the sample effect size (standardised mean difference) can be replicated or not.


\ac{ER} focuses on the effect of the $a$-run over the $b$-run, isolating it from other factors, but we may also want to account for absolute scores that are similar to the original experiment. Therefore, we propose to complement \ac{ER} with \acf{DeltaRI} and to plot \ac{ER} against \ac{DeltaRI} to visually interpret the replicability of the effects. We define \ac{DeltaRI} as follows\footnote{In Equation~\eqref{eq:deltari} we assume that both $\overline{M^{C}(b)}$ and $\overline{M^{C}(b')}$ are $> 0$. If these two values are equal to $0$, it means that the run score is equal to $0$ on each topic. Therefore, we can simply remove that run from the evaluation, as it is done for topics which do not have any relevant document.}:
\begin{equation}
     \text{RI} = \frac{\overline{M^{C}(a)} - \overline{M^{C}(b)}}{\overline{M^{C}(b)}}, \qquad \text{RI}' = \frac{\overline{M^{C}(a')} - \overline{M^{C}(b')}}{\overline{M^{C}(b')}} 
  \label{eq:deltari}
\end{equation}
where $\text{RI}$ and $\text{RI}'$ are the relative improvements for the original and replicated runs and $\overline{M^{C}(\cdot)}$ is the average score across topics. Now let \ac{DeltaRI} be
   $\Delta\text{RI}(\text{RI}, \text{RI}')  = \text{RI} - \text{RI}'$.
\ac{DeltaRI} ranges in $[-1, 1]$, $\Delta\text{RI} = 0$ means that the relative improvements are the same for the original and replicated runs; when $\Delta\text{RI} > 0$, the replicated relative improvement is smaller than the original relative improvement, and in case $\Delta\text{RI} < 0$, it is larger.
\ac{DeltaRI} can be used in combination with \ac{ER}, by plotting \ac{ER} ($x$-axis) against \ac{DeltaRI} ($y$-axis), as done in Figure~\ref{fig:rpl_er_ri}. If $ER = 1$ and $\Delta\text{RI} = 0$ both the effect and the relative improvements are replicated, therefore the closer a point to $(1, 0)$ the more successful the replication experiment. 
We can now divide the \ac{ER}-\ac{DeltaRI} plane in $4$ regions, corresponding to the $4$ quadrants of the cartesian plane:
\begin{itemize}[leftmargin=*]
    \item Region both $1$: \text{ER} $>0$ and \ac{DeltaRI} $>0$, the replication is somehow successful in terms of effect sizes, but not in terms of absolute scores;
    \item Region $2$: \text{ER} $<0$ and \ac{DeltaRI} $>0$, the replication is a failure both in terms of effect sizes and absolute scores;
    \item Region $3$: both \text{ER} $<0$ and \ac{DeltaRI} $<0$, the replication is a failure in terms of effect sizes, but not in terms of absolute scores;
    \item Region $4$: \text{ER} $>0$ and \ac{DeltaRI} $<0$, this means that the replication is somehow successful both in terms of effect sizes and absolute scores.
\end{itemize}
Therefore, the preferred region is Region $4$, with the condition that the best replicability runs are close to $(1,0)$.

\subsubsection*{Statistical Comparison}
We propose to compare the original and replicated runs in terms of their statistical difference: we run a two-tailed paired $t$-test between the scores of $r$ and $r'$ for each topic in $C$ with respect to an evaluation measure $M$. The $p$-value returned by the $t$-test informs on the extent to which $r$ is successfully replicated: the smaller the $p$ value, the stronger the evidence that $r$ and $r'$ are significantly different, thus $r'$ failed in replicating $r$.

Note that the $p$-value does not inform on the overall effect, i.e.~we may know that $r'$ failed to replicate $r$, but we cannot infer whether $r'$ performed better or worse than $r$.



\subsection{Reproducibility}
\label{subsec:reproducibility}

Differently from replicability, for reproducibility the original and reproduced runs are not obtained on the same collection (different documents and/or topic sets), thus the original run cannot be used for direct comparison with the reproduced run. As a consequence, Kendall's $\tau$, \ac{RBO}, and \ac{RMSE} in Section~\ref{subsec:replicability} cannot be applied to the reproducibility task. Therefore, hereinafter we focus on: (i) reproducing the \emph{overall effect} with \ac{ER}; (ii) comparing the original and reproduced runs with \emph{statistical tests}.

\subsubsection*{Overall Effect}
CENTRE@NTCIR~\cite{SakaiEtAl2019b} defines \ac{ER} for reproducibility as follows: given a pair of runs, $a$-run and $b$-run, where the $a$-run has been reported to outperform the $b$-run on a test collection $C$, can another research group reproduce the improvement on a different test collection $D$?
%
The original per-topic improvements are the same as in Eq.~\eqref{eq:per-topic-improvements}, while the reproduced per-topic improvements are defined as in Eq.~\eqref{eq:per-topic-improvements} by replacing $C$ with $D$. Therefore, the resulting \acf{ER}~\cite{SakaiEtAl2019b} is defined as follows:
\begin{equation}
\text{ER}(\Delta' M^{D}, \Delta M^{C}) = 
\frac{
\overline{\Delta' M^{D}}
}{
\overline{\Delta M^{C}}
}
=
\frac{
\frac{1}{n_{D}} \sum_{j=1}^{n_{D}} \Delta' M_{j}^{D}
}{
\frac{1}{n_{C}} \sum_{j=1}^{n_{C}} \Delta M_{j}^{C}
}
\label{eq:er_reproducibility}
\end{equation}
where $n_{D}$ is the number of topics in $D$. Assuming that the standard deviation of a measure $M$ is common across experiments, the above version of \ac{ER} is equivalent to the ratio of effect sizes (or standardised mean differences for the two-sample data case)~\cite{Sakai2018}; it can then be interpreted in a way similar to the \ac{ER} for replicability. Note that since we are considering the ratio of the mean improvements instead of the mean of the improvements ratio, Eq.~\eqref{eq:er_reproducibility} can be applied also when the number of topics in $C$ and $D$ is different. 

Similarly to the replicability case, \ac{ER} can be complemented with \ac{DeltaRI}, whose definition is the same of Eq.~\eqref{eq:deltari}, but $\text{RI}'$ is computed over the new collection $D$, instead of the original collection $C$. 
\ac{DeltaRI} has the same interpretation as in the replicability case, i.e.~to show if the improvement in terms of relative scores in the reproduced experiment are similar to the original experiment.

\subsubsection*{Statistical Comparison}
  
With a t-test, we can also handle the case when the original and the reproduced experiments are based on different datasets. In this case, we need to perform a two-tailed unpaired t-test to account, for the different subjects used in the comparison. 

The unpaired t-test assumes equal variance and this is likely to not happen when, e.g., you have two different sets of topics in the two datasets. However, the unpaired t-test is known to be robust to such violations and \citet{Sakai2016b} has shown that Welch's t-test, which assumes unequal variance, may be less reliable when the sample sizes differ substantially and the larger sample has a substantially larger variance.




\section{Dataset}
\label{sec:dataset}

To evaluate the measures in Section~\ref{sec:framework}, we need a re\-pro\-du\-ci\-bi\-li\-ty-oriented dataset and, to the best of our knowledge, this is the first attempt to construct such a dataset. The use case behind our dataset is that of a researcher who tries to replicate the methods described in a paper and who also tries to reproduce those results on a different collection; the researcher uses the presented measures as a guidance to select the best replicated/reproduced run and understand when reproduced is reproduced.
Therefore, to cover both replicability and reproducibility, the dataset should contain both a baseline and an advanced run. Furthermore, the dataset should contain runs with different ``quality levels'', roughly meant as being more or less ``close'' to the orginal run, to mimic the different attempts of a researcher to get closer and closer to the original run.

We reimplement \texttt{WCrobust04} and \texttt{WCrobust0405}, two runs submitted by~\citet{DBLP:conf/trec/GrossmanC17} to the TREC 2017 Common Core track~\cite{AllanEtAl2017b}.  \texttt{WCrobust04} and \texttt{WCrobust0405} rank documents by routing using profiles~\cite{robertson_callan}. In particular, \citeauthor{DBLP:conf/trec/GrossmanC17} extract relevance feedback from a \emph{training corpus}, train a logistic regression classifier with tfidf-features of relevant documents to a topic, and rank documents of a \emph{target corpus} by their probability of being relevant to the same topic. The baseline run and the advanced run differ by the training data used for the classifier -- one single corpus for \texttt{WCrobust04}, two corpora for \texttt{WCrobust0405}. We \emph{replicate} runs using The New York Times Corpus, our target corpus; we \emph{reproduce} runs using Washington Post Corpus.
It is a requirement that all test collections, i.e., those used for training as well as the target collection, share at least some of the same topics. Our replicated runs cover $50$ topics, whereas the reproduced runs cover $25$ topics. 
Full details on the implementation can be found in ~\cite{DBLP:conf/clef/BreuerS19} and in the public repository\footnote{https://github.com/irgroup/sigir2020-measure-reproducibility}~\cite{timo_breuer_2020_3856042}, which also contains the full dataset, consisting of $200$ runs.

To generate replicated and reproduced runs, we systematically change a set of parameters and derive $4$ \emph{constellations} consisting of $20$ runs each, for a total of $80$ runs ($40$ runs for replicability and $40$ runs for reproducibility)\footnote{An alternative to our approach could be to artificially alter one or more existing runs by swapping and/or changing retrieved documents or, even, to generate artificial runs fully from scratch. However, these artificial runs would have had no connection with the principled way in which a researcher actually proceeds when trying to reproduce an experiment and with her/his need to get orientation during this process. As a result, an artificially constructed dataset would lack any clear use case behind it.}. We call them constellations because, by gradually changing the way in which training features are generated and the classifier is parameterized, we obtain sets of runs which are further and further away from the original run in a somehow controlled way and, in Section~\ref{subsec:performance_analysis}, we will exploit this regularity to validate the behaviour of our measures. The $4$ constellations are:

\begin{itemize}[leftmargin=*]
\item \texttt{rpl\_wcr04\_tf}\footnote{The exemplified denotation applies to the replicated baseline run. The advanced and reproduced runs are denotated according to this scheme.}: These runs incrementally reduce the vocabulary size by limiting it with the help of a threshold. Only those tfidf-features with a term frequency above the specified threshold are considered.
\item \texttt{rpl\_wcr04\_df}: Alternatively, the vocabulary size can be reduced by the document frequency. In this case, only terms with a document frequency below a specified maximum are considered. This means common terms included in many documents are excluded.
\item \texttt{rpl\_wcr04\_tol}: Starting from a default parametrization of the classifier, we increase the tolerance of the stopping criterion. Thus, the training is more likely to end earlier at the cost of accuracy.
\item \texttt{rpl\_wcr04\_C}: Comparable to the previous constellation, we start from a default parametrization and vary the $\ell^{2}$-regulari\-zation strength towards poorer accuracy.
\end{itemize}
These constellations are examples of typical implementation details that might be considered as part of the principled way of a reproducibility study. If no information on the exact configuration is given, the researcher has to guess reasonable values for these parameters and thus to produce different runs.

Beside the above constellations, the dataset includes runs with several other configurations obtained by excluding pre-processing steps, varying the generation of the vocabulary, applying different tfidf-formulations, using n-grams with varying lengths, or implementing a support-vector machine as the classifier. This additional constellation, containing $120$ runs ($60$ runs for replicability and $60$ runs for reproducibility), consists of runs which vary in a sharper and less regular way. In Section~\ref{subsec:correlation_analysis}, we will exploit this constellation together with the previous ones to conduct a correlation analysis and understand how our proposed measures are related in a more general case.

\section{Experimental Evaluation}
\label{sec:experiments_new}

We evaluate our measures in two ways. Firstly, using the first $4$ ``regular'' constellations described in Section~\ref{sec:dataset}, we check that our measures behave as expected in these known cases, roughly speaking we check that they tend to increase/decrease as expected. Secondly, using all the constellations described in Section~\ref{sec:dataset},  we check that our measures actually provide different viewpoints on replicability/reproducibilty by conducting a correlation analysis. To this end, as usual, we compute Kendall's $\tau$ correlation\footnote{We choose Kendall's $\tau$ because, differently from Spearman's correlation coefficient, it can handle ties and it also has better statistical properties than Pearson's correlation coefficient~\cite{CrouxAndDehon2010}. We did not consider AP correlation~\cite{YilmazEtAl2008} since, as shown in~\cite{Ferro2017}, it ranks measures in the same way as Kendall's $\tau$.} among the rankings of runs produced by each of our measures. Whenever the correlation between two measures is very high, we can report just one measure, since the other will likely represent redundant information~\cite{WebberEtAl2008}; furthermore, as suggested by~\citet{Voorhees1998}, we consider two measures equivalent if their correlation is greater than $0.9$, and noticeably different if Kendall's $\tau$ is below $0.8$.

As effectiveness measures used with \ac{ARP}, \ac{RMSE} and \ac{ER}, we select \ac{AP} and \ac{nDCG} with cut-off $1000$ and P@10. Even if P@10 might be redundant~\cite{WebberEtAl2008}, we want to investigate whether it is easier to replicate/reproduce an experiment with a set-based measure. RBO is computed with $\phi = 0.8$. Even if~\citet{WebberEtAl2010} instantiate \ac{RBO} with $\phi \geq 0.9$, we exploit a lower $\phi$. Inspired by the analysis for \ac{RBP} in~\citet{FerranteEtAl2015b-nf}, we select a lower $\phi$ to consider a less top-heavy measure, since for replicability we do not want to replicate just the top rank positions.

\subsection{Validation of Measures}
\label{subsec:performance_analysis}

\subsubsection*{Case Study: Replicability}

\begin{table*}[tb]
\caption{Replicability results for \texttt{WCrobust04}: \ac{ARP}, rank correlations, \ac{RMSE}, and $p$-values returned by the paired $t$-test.}
\label{tab:replicability_04}
\resizebox{0.8\textwidth}{!}{%
\begin{tabular}{@{}l|ccc|cc|ccc|ccc@{}}
\toprule
    & \multicolumn{3}{c|}{\ac{ARP}} & \multicolumn{2}{c|}{Correlation} & \multicolumn{3}{c|}{RMSE} & \multicolumn{3}{c}{$p$-value} \\
    run & P@10  & AP & nDCG & $\tau$ & \ac{RBO} & P@10 & AP & nDCG & P@10 & AP & nDCG \\ 
    \midrule
    \texttt{WCrobust04}	& $0.6460$	& $0.3711$	& $0.6371$	& $1$	& $1$	& $0$	& $0$	& $0$	& $1$	& $1$	& $1$ \\
    \midrule
    \texttt{rpl\_wcr04\_tf\_1}	& $0.6920$	& $0.3646$	& $0.6172$	& $0.0117$	& $0.5448$	& $0.2035$	& $0.0755$	& $0.0796$	& $0.110$	& $0.551$	& $0.077$ \\
    \texttt{rpl\_wcr04\_tf\_2}	& $0.6900$	& $0.3624$	& $0.6177$	& $0.0096$	& $0.5090$	& $0.2088$	& $0.0799$	& $0.0810$	& $0.137$	& $0.445$	& $0.090$ \\
    \texttt{rpl\_wcr04\_tf\_3}	& $0.6820$	& $0.3420$	& $0.6011$	& $0.0076$	& $0.4372$	& $0.2375$	& $0.1083$	& $0.0971$	& $0.288$	& $0.056$	& $0.007$ \\
    \texttt{rpl\_wcr04\_tf\_4}	& $0.6680$	& $0.3106$	& $0.5711$	& $0.0037$	& $0.3626$	& $0.2534$	& $0.1341$	& $0.1226$	& $0.544$	& $9E{-}04$	& $4E{-}05$ \\
    \texttt{rpl\_wcr04\_tf\_5}	& $0.6220$	& $0.2806$	& $0.5365$	& $0.0064$	& $0.2878$	& $0.2993$	& $0.1604$	& $0.1777$	& $0.575$	& $1E{-}05$	& $1E{-}05$ \\
    \midrule
    \texttt{rpl\_wcr04\_df\_1}	& $0.6700$	& $0.3569$	& $0.6145$	& $0.0078$	& $0.5636$	& $0.2000$	& $0.0748$	& $0.0742$	& $0.401$	& $0.181$	& $0.029$ \\
    \texttt{rpl\_wcr04\_df\_2}	& $0.6560$	& $0.3425$	& $0.6039$	& $0.0073$	& $0.5455$	& $0.1772$	& $0.0779$	& $0.0802$	& $0.694$	& $0.008$	& $0.002$ \\
    \texttt{rpl\_wcr04\_df\_3}	& $0.6020$	& $0.3049$	& $0.5692$	& $0.0072$	& $0.5217$	& $0.1649$	& $0.1078$	& $0.1210$	& $0.058$	& $1E{-}06$	& $1E{-}05$ \\
    \texttt{rpl\_wcr04\_df\_4}	& $0.5220$	& $0.2519$	& $0.5058$	& $0.0048$	& $0.4467$	& $0.2098$	& $0.1695$	& $0.1987$	& $4E{-}06$	& $8E{-}09$	& $1E{-}07$ \\
    \texttt{rpl\_wcr04\_df\_5}	& $0.4480$	& $0.2121$	& $0.4512$	& $0.0019$	& $0.3532$	& $0.3102$	& $0.2053$	& $0.2572$	& $4E{-}07$	& $2E{-}11$	& $2E{-}09$ \\
    \midrule
    \texttt{rpl\_wcr04\_tol\_1}	& $0.6700$	& $0.3479$	& $0.5992$	& $0.0033$	& $0.5504$	& $0.2010$	& $0.0783$	& $0.0928$	& $0.403$	& $0.035$	& $0.002$ \\
    \texttt{rpl\_wcr04\_tol\_2}	& $0.5680$	& $0.2877$	& $0.4901$	& $0.0061$	& $0.4568$	& $0.3216$	& $0.1868$	& $0.2931$	& $0.086$	& $0.001$	& $1E{-}04$ \\
    \texttt{rpl\_wcr04\_tol\_3}	& $0.3700$	& $0.1812$	& $0.3269$	& $0.0066$	& $0.2897$	& $0.4762$	& $0.2937$	& $0.4387$	& $8E{-}06$	& $2E{-}07$	& $6E{-}09$ \\
    \texttt{rpl\_wcr04\_tol\_4}	& $0.2180$	& $0.0903$	& $0.1728$	& $0.0066$	& $0.1621$	& $0.5488$	& $0.3512$	& $0.5382$	& $1E{-}11$	& $1E{-}12$	& $4E{-}16$ \\
    \texttt{rpl\_wcr04\_tol\_5}	& $0.0700$	& $0.0088$	& $0.0379$	& $0.0012$	& $0.0518$	& $0.6437$	& $0.4028$	& $0.6228$	& $8E{-}19$	& $3E{-}19$	& $2E{-}29$ \\
    \midrule
    \texttt{rpl\_wcr04\_C\_1}	& $0.7020$	& $0.3671$	& $0.6191$	& $0.0039$	& $0.5847$	& $0.1744$	& $0.0631$	& $0.0640$	& $0.021$	& $0.656$	& $0.046$ \\
    \texttt{rpl\_wcr04\_C\_2}	& $0.6960$	& $0.3717$	& $0.6244$	& $0.0021$	& $0.5907$	& $0.1772$	& $0.0610$	& $0.0606$	& $0.044$	& $0.945$	& $0.142$ \\
    \texttt{rpl\_wcr04\_C\_3}	& $0.6840$	& $0.3532$	& $0.6093$	& $0.0096$	& $0.5607$	& $0.2168$	& $0.0833$	& $0.0850$	& $0.218$	& $0.130$	& $0.019$ \\
    \texttt{rpl\_wcr04\_C\_4}	& $0.6240$	& $0.3168$	& $0.5761$	& $0.0073$	& $0.4595$	& $0.2249$	& $0.1144$	& $0.1194$	& $0.494$	& $4E{-}04$	& $1E{-}04$ \\
    \texttt{rpl\_wcr04\_C\_5}	& $0.6140$	& $0.3085$	& $0.5689$	& $0.0068$	& $0.4483$	& $0.2315$	& $0.1192$	& $0.1248$	& $0.333$	& $7E{-}05$	& $3E{-}05$ \\
    \bottomrule
\end{tabular}%
}
\end{table*}

\begin{figure}[tb]
\begin{subfigure}{.49\linewidth}
\centering
\includegraphics[width=\textwidth]{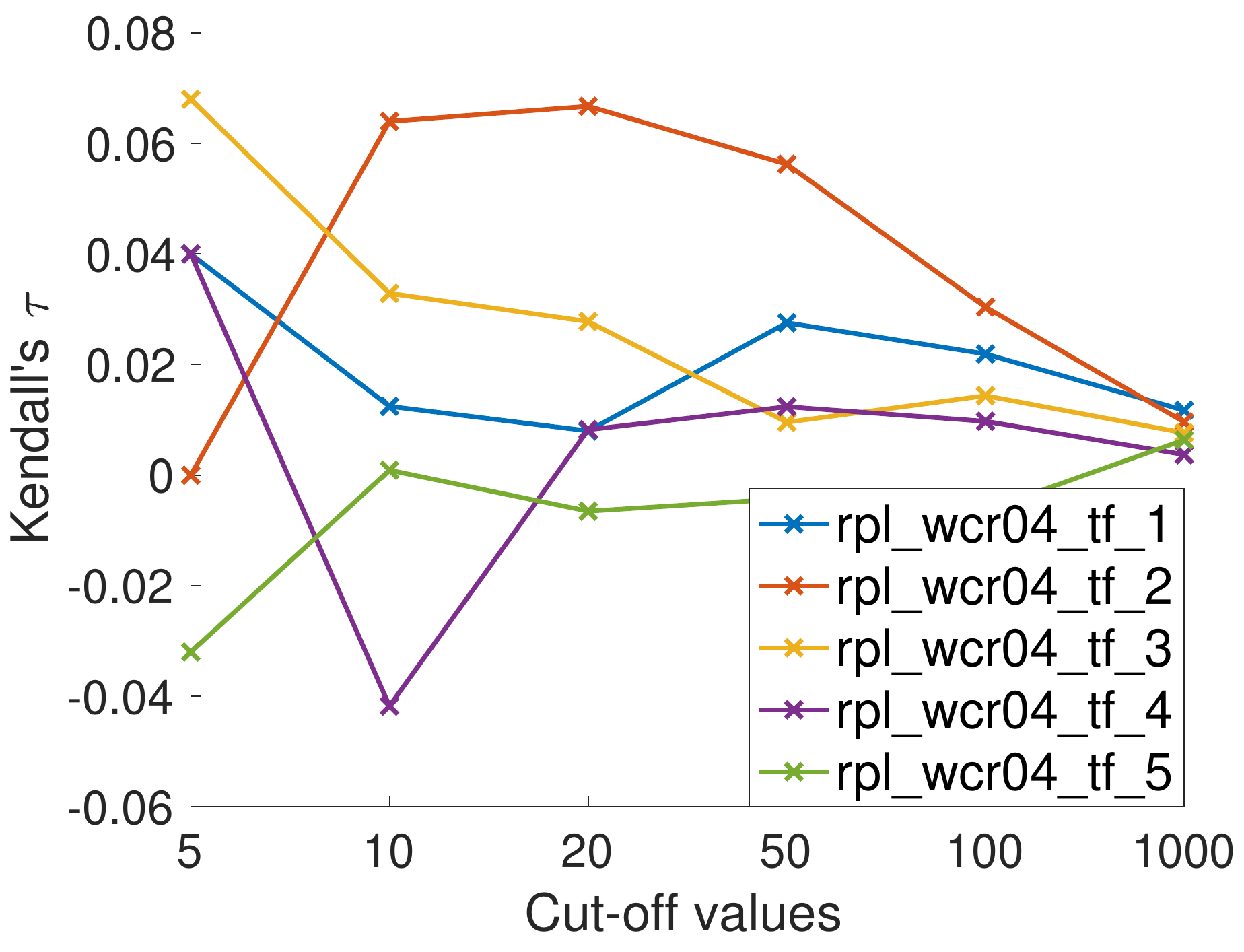}
\caption{Kendall's $\tau$ \texttt{WCrobust04}.}
\label{fig:tau_cutoff_04}
\end{subfigure}%
\hfill
\centering
\begin{subfigure}{.49\linewidth}
\includegraphics[width=\textwidth]{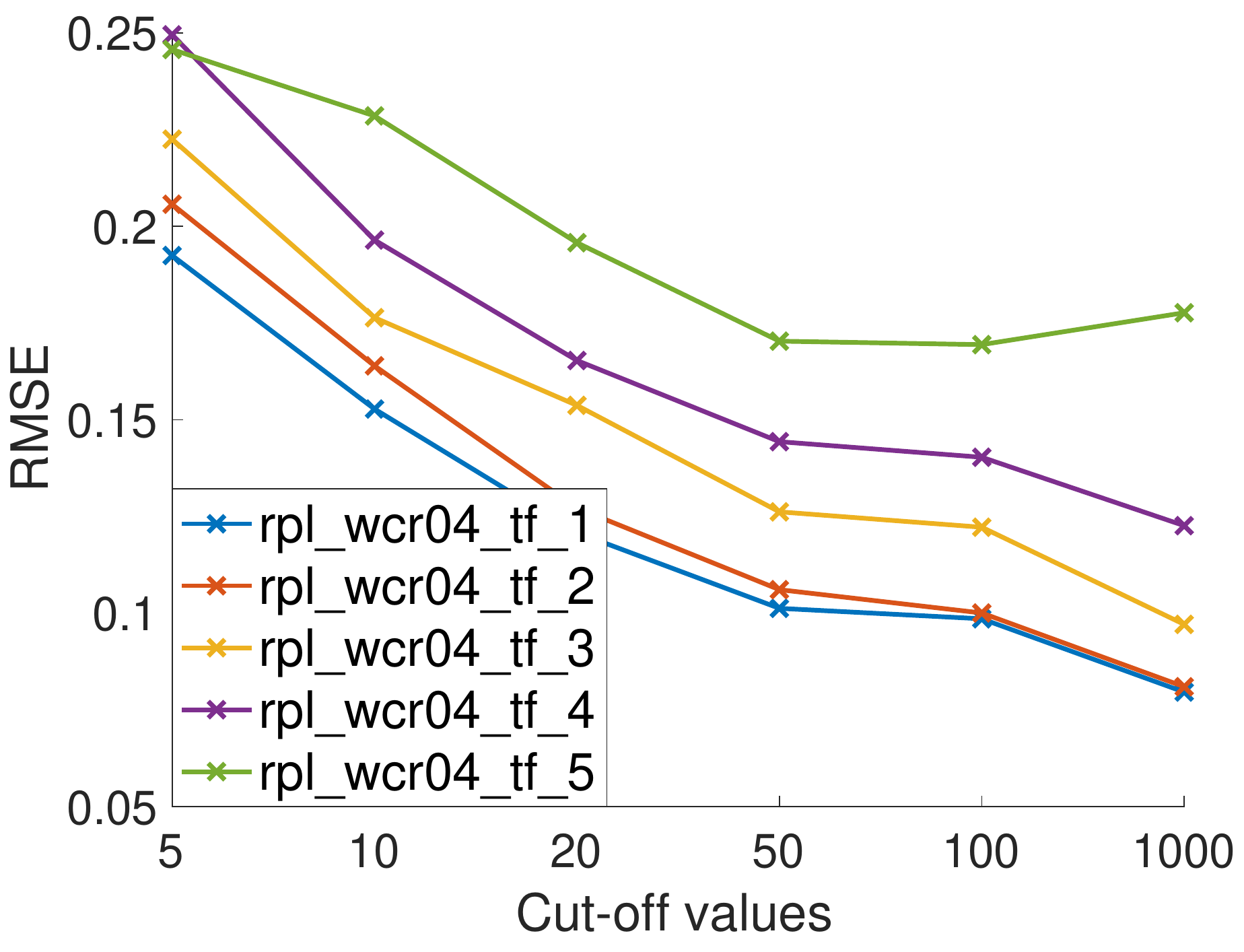}
\caption{\ac{RMSE} \texttt{WCrobust04}.}
\label{fig:rmse_cutoff_04}
\end{subfigure}%
\caption{Kendall's $\tau$ and \ac{RMSE} with \ac{nDCG} computed at different cut-offs for \texttt{WCrobust04}.}
\label{fig:rmse_tau_cutoff}
\end{figure}

\begin{figure}[tb]
\centering
\begin{subfigure}{.49\linewidth}
\includegraphics[width=\textwidth]{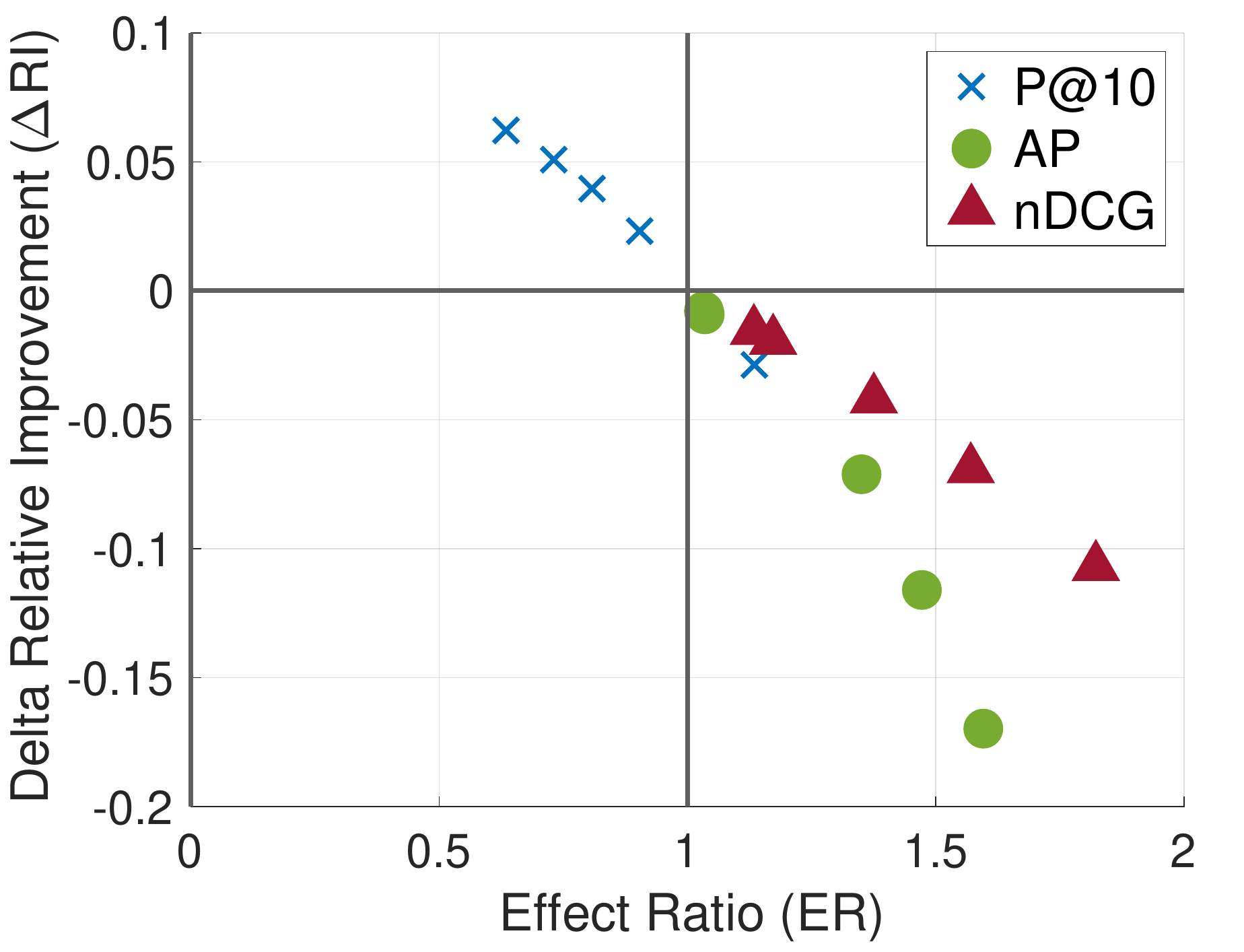}
\caption{\texttt{rpl\_tf} runs.}
\label{fig:er_ri_tf}
\end{subfigure}%
\hfill
\begin{subfigure}{.49\linewidth}
\includegraphics[width=\textwidth]{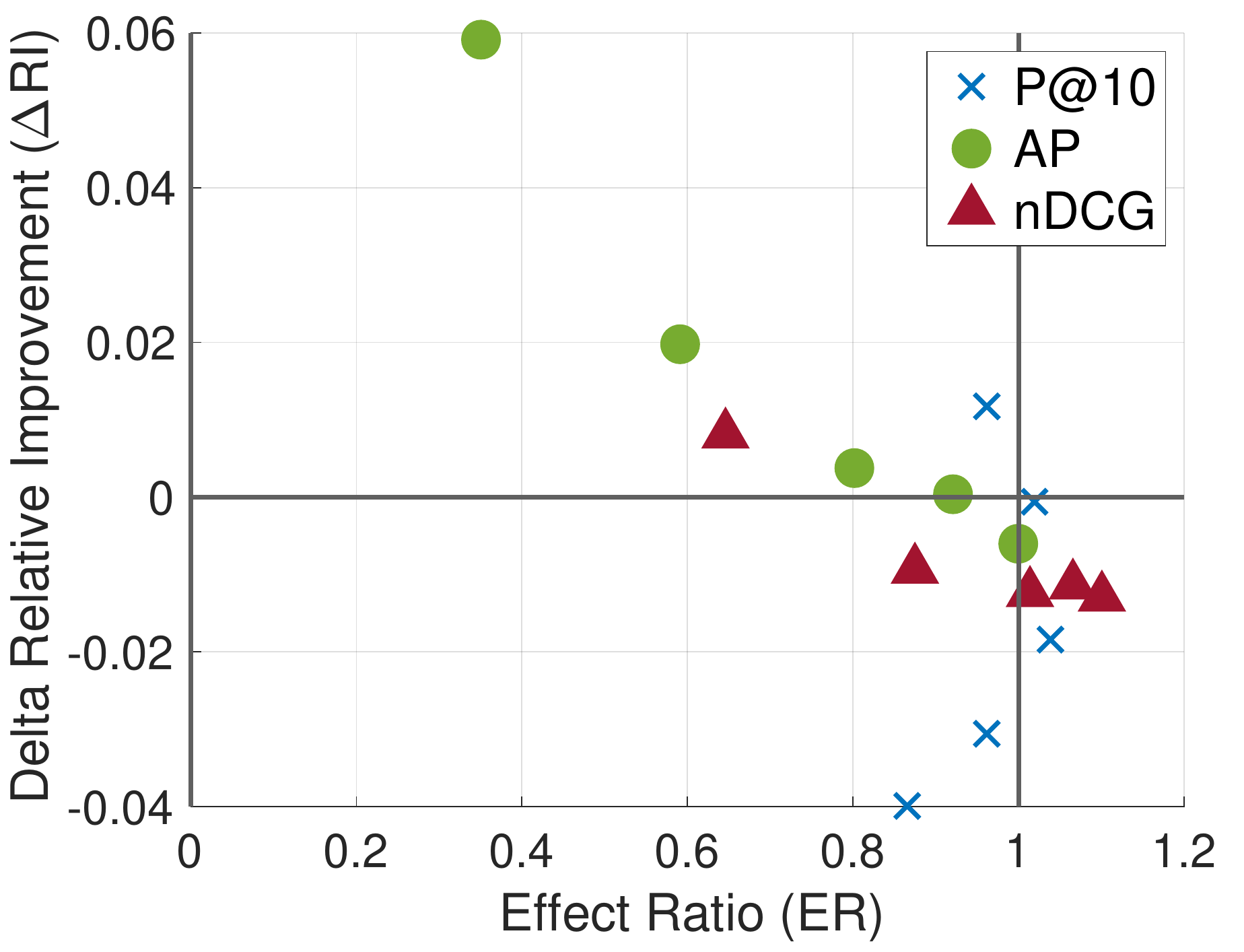}
\caption{\texttt{rpl\_df} runs.}
\label{fig:er_ri_df}
\end{subfigure}%
\caption{Replicability: \ac{ER} on the $x$-axis against \ac{DeltaRI} on the $y$-axis.}
\label{fig:rpl_er_ri}
\end{figure}

Table~\ref{tab:replicability_04} reports the retrieval performance for the baseline $b$-run \texttt{WCrobust04} and the replicability measures: Kendall's $\tau$, \ac{RBO}, \ac{RMSE}, and the $p$-values returned by the paired $t$-test. The corresponding table for \texttt{WCrobust0405} reports similar results and is included in an online appendix~\footnote{\url{https://github.com/irgroup/sigir2020-measure-reproducibility/tree/master/appendix}}. 
We report \ac{ER} in Table~\ref{tab:replicability_er} and plot \ac{ER} against DeltaRI in Figure~\ref{fig:rpl_er_ri}, additional \ac{ER}-DeltaRI plots are included in the online appendix.

In Table~~\ref{tab:replicability_04}, low values for Kendall's $\tau$ and \ac{RBO} highlights how hard it is to accurately replicate a run at ranking level. Replicability runs achieve higher \ac{RBO} scores than Kendall's $\tau$, showing that \ac{RBO} is somehow less strict.

\ac{RMSE} increases almost consistently when the difference between \ac{ARP} scores of the original and replicated runs decreases. In general, \ac{RMSE} values of P@10 are larger compared to those of \ac{AP} and \ac{nDCG}, due to P@10 having naturally higher variance (since it also considers a lower cut-off). For the constellation \texttt{rpl\_wcr04\_tf} and \texttt{rpl\_wcr04\_C}, \ac{RMSE} with P@10 increases, even if the difference between \ac{ARP} scores decreases. As pointed out in Section~\ref{subsec:replicability}, this is due to \ac{RMSE} which penalizes large errors. 
On the other hand, \ac{RMSE} decreases almost consistently as the cut-off value increases, as shown in Figure~\ref{fig:rmse_cutoff_04}. As expected, if we consider the whole ranking, the replicability runs retrieve more relevant documents and thus achieve better \ac{RMSE} scores. 

As a general observation, it is easier to replicate a run in terms of \ac{RMSE} rather than Kendall's $\tau$ or \ac{RBO}. This is further corroborated by the correlation results in Table~\ref{tab:replicability_correlation}, which shows low correlation between \ac{RMSE} and Kendall's $\tau$. Therefore, even if the original and the replicated runs place documents with the same relevance labels in the same rank positions, those documents are not the same, as shown in Figure~\ref{fig:tau_cutoff_04}, where Kendall's $\tau$ is computed at different cut-offs. This does not affect the system performance, but it might affect the user experience, which can be completely different.

For the paired $t$-test, as the difference in \ac{ARP} decreases, $p$-value increases, showing that the runs are more similar. This is further validated by high correlation results reported in Table~\ref{tab:replicability_correlation} between \ac{ARP} and $p$-values. Recall that the numerator of the $t$-value is basically computing the difference in \ac{ARP} scores, thus explaining the consistency of these results. 

For \texttt{rpl\_wcr04\_tf} and \texttt{rpl\_wcr04\_C}, \ac{RMSE} and $p$-values are not consistent: \ac{RMSE} increases, thus the error increases, but $p$-values also increase, thus the runs are considered more similar. As aforementioned, this happens because \ac{RMSE} penalizes large errors per topic, while the $t$-statistic is tightly related to \ac{ARP} scores.

\begin{table}[tb] \setlength{\tabcolsep}{3.5pt}
\caption{ER results for replicability and reproducibility: the $a$-run
  is \texttt{WCrobust0405} on TREC Common Core 2017; the
  $b$-run is \texttt{WCrobust04}, for replicability on TREC Common
  Core 2017, for reproducibility on TREC Common Core 2018.}
\label{tab:replicability_er}
\begin{tabular}{@{}l|ccc||ccc@{}}
  \toprule
  &\multicolumn{3}{c||}{replicability}&\multicolumn{3}{c}{reproducibility}\\
    run & P@10 & \ac{AP} & \ac{nDCG} &                                            P@10                 & \ac{AP} & \ac{nDCG} \\ 
    \midrule                                                                                                                   
    \texttt{rpl\_tf\_1}	& $0.8077$	& $1.0330$	& $1.1724$ &               $1.1923$	& $1.2724$	& $2.0299$ \\
    \texttt{rpl\_tf\_2}	& $0.7308$	& $1.0347$	& $1.1336$ &               $0.9615$	& $1.3195$	& $2.2139$ \\
    \texttt{rpl\_tf\_3}	& $0.9038$	& $1.3503$	& $1.3751$ &               $1.5000$	& $1.5616$	& $2.5365$ \\
    \texttt{rpl\_tf\_4}	& $0.6346$	& $1.4719$	& $1.5703$ &               $1.4231$	& $1.9493$	& $2.9317$ \\
    \texttt{rpl\_tf\_5}	& $1.1346$	& $1.5955$	& $1.8221$ &               $1.5385$	& $1.7010$	& $3.0569$ \\
    \midrule                                                                                                                                         
    \texttt{rpl\_df\_1}	& $0.9615$	& $0.9995$	& $1.1006$ &               $0.4615$	& $0.7033$	& $0.9547$ \\
    \texttt{rpl\_df\_2}	& $1.0192$	& $0.9207$	& $1.0656$ &               $0.4231$	& $0.4934$	& $0.6586$ \\
    \texttt{rpl\_df\_3}	& $1.0385$	& $0.8016$	& $1.0137$ &               $0.1923$	& $0.5429$	& $1.0607$ \\
    \texttt{rpl\_df\_4}	& $0.9615$	& $0.5911$	& $0.8747$ &               $0.3846$	& $0.5136$	& $0.8333$ \\
    \texttt{rpl\_df\_5}	& $0.8654$	& $0.3506$	& $0.6459$ &               $0.3846$	& $0.4857$	& $0.7260$ \\
    \midrule                                                                                                                                         
    \texttt{rpl\_tol\_1}	& $1.0769$	& $1.2013$	& $1.3455$ &      $0.5769$	& $0.6574$	& $0.8780$ \\
    \texttt{rpl\_tol\_2}	& $1.3269$	& $1.4946$	& $1.9290$ &      $0.8077$	& $0.5194$	& $0.8577$ \\
    \texttt{rpl\_tol\_3}	& $1.8654$	& $2.1485$	& $2.8496$ &      $2.0000$	& $1.4524$	& $2.9193$ \\
    \texttt{rpl\_tol\_4}	& $2.0962$	& $2.2425$	& $3.3213$ &      $2.3846$	& $2.1242$	& $3.9092$ \\
    \texttt{rpl\_tol\_5}	& $1.2500$	& $1.0469$	& $1.8504$ &      $0.2692$	& $0.1116$	& $0.5595$ \\
    \midrule                                                                                                                   
    \texttt{rpl\_C\_1}	& $0.6346$	& $0.6300$	& $0.8901$ &               $2.1538$	& $1.8877$	& $3.7777$ \\
    \texttt{rpl\_C\_2}	& $0.8077$	& $0.7361$	& $0.9240$ &               $2.2308$	& $1.9644$	& $3.8621$ \\
    \texttt{rpl\_C\_3}	& $0.8654$	& $1.1195$	& $1.2092$ &               $2.3846$	& $2.2743$	& $4.2783$ \\
    \texttt{rpl\_C\_4}	& $0.9231$	& $1.1642$	& $1.2911$ &               $0.6538$	& $0.7316$	& $1.0403$ \\
    \texttt{rpl\_C\_5}	& $0.8846$	& $1.1214$	& $1.2542$ &               $0.5769$	& $0.6915$	& $0.9741$ \\
    \bottomrule
\end{tabular}
\end{table}

Table~\ref{tab:replicability_er} (left) reports \ac{ER} scores for replicability runs. \texttt{WCrobust\_04} is the baseline $b$-run, while \texttt{WCrobust\_0405} is the advanced $a$-run, both of them on TREC Common Core 2017. Recall that, for \ac{ER}, the closer the score to $1$, the more successful the replication. 

\ac{ER} behaves as expected: when the quality of the replicated runs deteriorates, \ac{ER} scores tend to move further from $1$. As for \ac{RMSE}, we can observe that the extent of success for the replication experiments depends on the effectiveness measure. Thus, the best practice is to consider multiple effectiveness measures.

Note that, for the constellations of runs \texttt{rpl\_wcr04\_tf} and \texttt{rpl\_\-wcr04\_\-C}, there is no agreement among the best replication experiment when different effectiveness measures are considered. This trend is similar to the one observed with \ac{RMSE}, $p$-values and delta in \ac{ARP}. For example, for \ac{ER} with P@10, the best replicability runs are \texttt{rpl\_wcr04\_tf3} and \texttt{rpl\_}\texttt{wcr0405\_tf3} but \ac{ER} scores are not stable, while for \ac{AP} and \ac{nDCG}, \ac{ER} values tends to move further from $1$, as we deteriorate the replicability runs. Again, this is due to the high variance of P@10.

Figure~\ref{fig:rpl_er_ri} illustrates \ac{ER} scores against \ac{DeltaRI} for $2$ constellations in Table~\ref{tab:replicability_er} and the other constellations are included in the online appendix. Recall that in Figure~\ref{fig:rpl_er_ri}, the closer a point to the reference $(1, 0)$, the better the replication experiment, both in terms of effect sizes and absolute differences.

The \ac{ER}-\ac{DeltaRI} plot, can be used as a visual tool to guide researcher on the exploration of the ``space of replicability'' runs. For example, in Figure~~\ref{fig:er_ri_tf}, for \ac{AP} and \ac{nDCG} the point $(1, 0)$ is reached from Region $4$, which is somehow the preferred region, since it corresponds to successful replication both in terms of effect sizes and relative improvements. Conversely, in Figure~\ref{fig:er_ri_df}, it is clear that for \ac{AP} the point $(1, 0)$ is reached from Region $1$, which corresponds to somehow a successful replication in terms of effect sizes, but not in terms of relative improvements.

\subsubsection*{Case Study: Reproducibility}

For reproducibility, Table~\ref{tab:reproducibility_p_value} reports \ac{ARP} and $p$-values in terms of P@10, \ac{AP}, and \ac{nDCG}, for the runs reproducing \texttt{WCrobust04} on TREC Common Core 2018. The corresponding table for \texttt{WCrobust0405} is included in the online appendix. Note that, in this case we do not have the original run scores, so we cannot directly compare \ac{ARP} values. This represents the main challenge when evaluating reproducibility runs.

From $p$-values in Table~\ref{tab:reproducibility_p_value}, we can conclude that all the reproducibility runs are statistically significantly different from the original run, being the highest $p$-value just $0.005$. Therefore, it seems that none of the runs successfully reproduced the original run. 

However, this is likely due to the two collections being too different, which in turn makes the scores distribution also different. Consequently the $t$-test considers all the distributions as significantly different. To validate this hypothesis, we carried out an unpaired $t$-test between pairs of replicability and reproducibility runs in the $4$ different constellations. This means that each pair of runs is generated by the same system on two different collections. The $p$-values for this experiment are reported only in the online appendix. Again, the majority of the runs are considered statistically differerent, except for a few cases for \texttt{rpl\_wcr04\_df} and \texttt{rpl\_wcr04\_tol}, which exhibit higher $p$-values also in Table~\ref{tab:reproducibility_p_value}. This shows that, depending on the collections, the unpaired $t$-test can fail in correctly detecting reproduced runs.

Table~\ref{tab:replicability_er} (right) reports \ac{ER} scores for replicability runs. At a first sight, we can see that \ac{ER} scores are much lower (close to $0$) or much higher ($\gg 1$) than for the replicability case. If it is hard to perfectly replicate an experiment, it is even harder to perfectly reproduce it.

\begin{figure}[tb]
\centering
\begin{subfigure}{.49\linewidth}
\includegraphics[width=\textwidth]{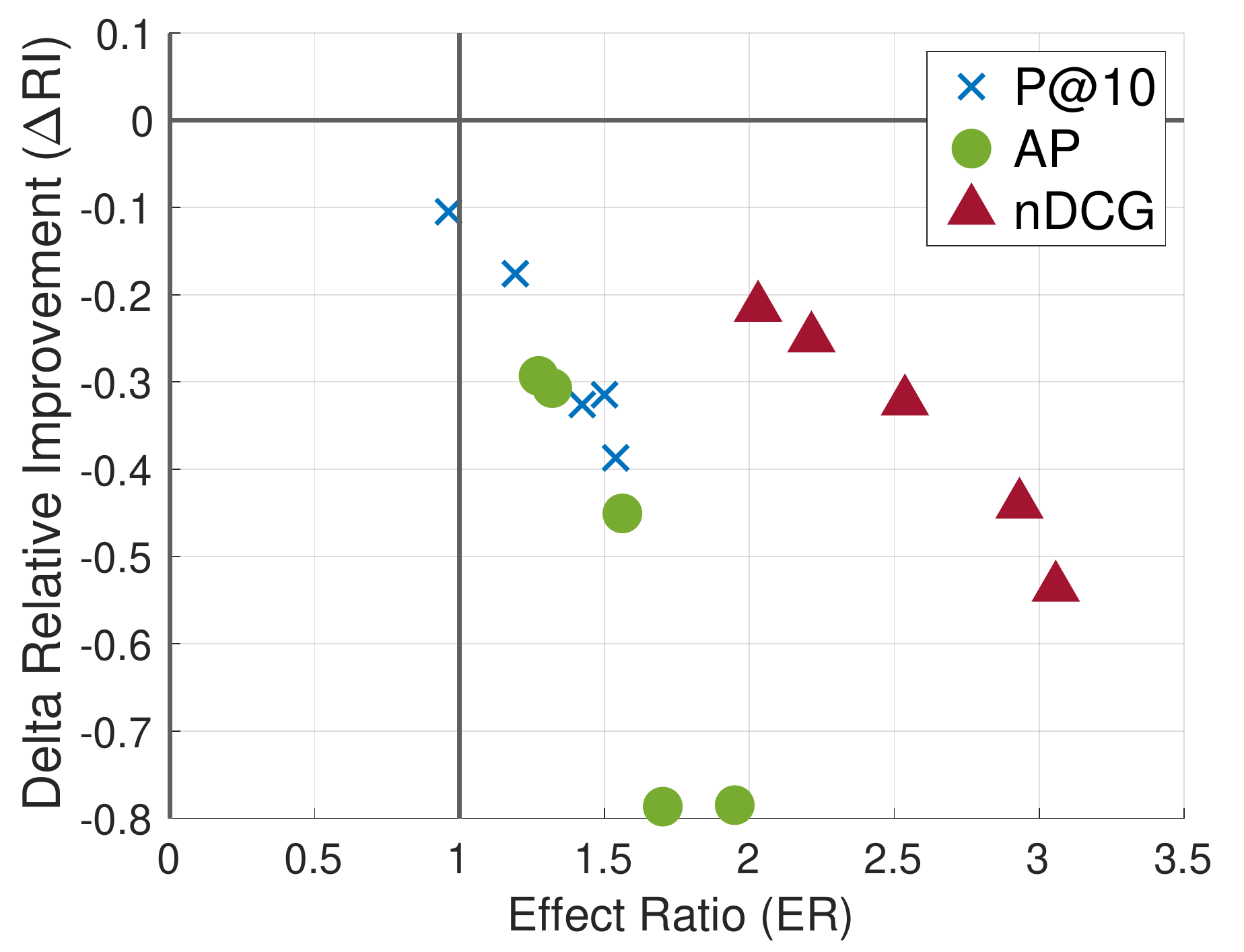}
\caption{\texttt{rpd\_tf} runs.}
\label{fig:rpd_er_ri_tf}
\end{subfigure}%
\hfill
\begin{subfigure}{.49\linewidth}
\includegraphics[width=\textwidth]{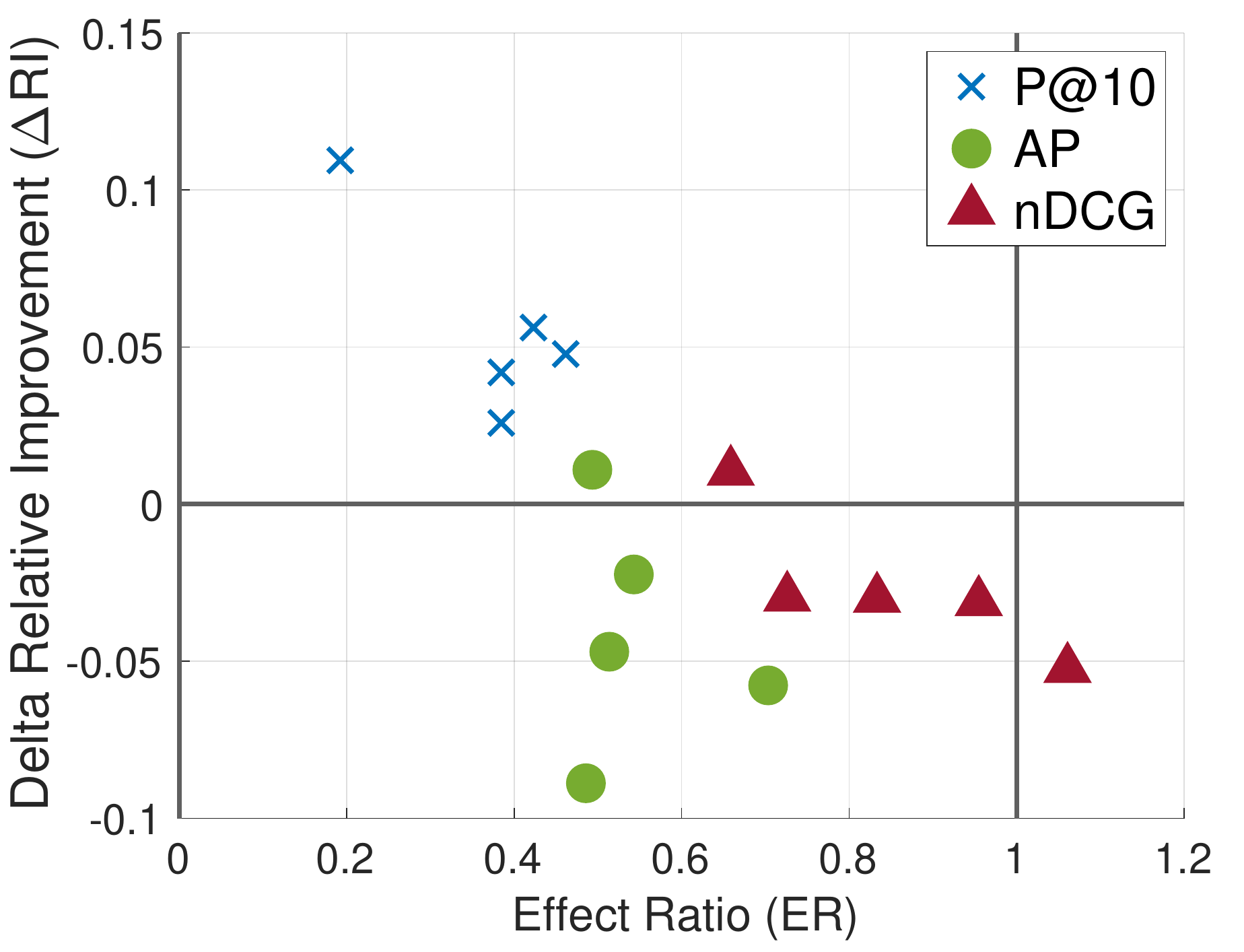}
\caption{\texttt{rpd\_df} runs.}
\label{fig:rpd_er_ri_df}
\end{subfigure}%
\caption{Reproducibility: \ac{ER} on the $x$-axis against \ac{DeltaRI} on the $y$-axis.}
\label{fig:rpd_er_ri}
\end{figure}

This is illustrated in the \ac{ER}-\ac{DeltaRI} plot in Figure~\ref{fig:rpd_er_ri}. In Figure~\ref{fig:rpd_er_ri_tf} the majority of the points are far from the best reproduction $(1, 0)$, even if they are in region $4$. In Figure~\ref{fig:rpd_er_ri_df} just one point is in the preferred region $4$, while many points are in region $2$, that is failure both in reproducing the effect size and the relative improvement. 

\begin{table}[tb]
\caption{Reproducibility: \ac{ARP} and $p$-value (unpaired $t$-test), for \texttt{WCrobust04}. The original runs are on TREC Common Core 2017, and reproduced runs on TREC Common Core 2018.}
\label{tab:reproducibility_p_value}
\resizebox{\columnwidth}{!}{%
\begin{tabular}{@{}l|ccc|ccc@{}}
\toprule
    & \multicolumn{3}{c|}{\ac{ARP}} & \multicolumn{3}{c}{$p$-value} \\
    run & P@10 & \ac{AP} & \ac{nDCG} & P@10 & \ac{AP} & \ac{nDCG} \\ 
    \midrule
    \texttt{rpd\_tf\_1}	& $0.3680$	& $0.1619$	& $0.3876$	& $7E{-}04$	& $6E{-}06$	& $6E{-}06$	 \\
    \texttt{rpd\_tf\_2}	& $0.3760$	& $0.1628$	& $0.3793$	& $9E{-}04$	& $8E{-}06$	& $4E{-}06$	 \\
    \texttt{rpd\_tf\_3}	& $0.3280$	& $0.1468$	& $0.3587$	& $8E{-}05$	& $1E{-}06$	& $8E{-}07$	 \\
    \texttt{rpd\_tf\_4}	& $0.3040$	& $0.1180$	& $0.3225$	& $2E{-}05$	& $3E{-}08$	& $1E{-}08$	 \\
    \texttt{rpd\_tf\_5}	& $0.2920$	& $0.1027$	& $0.2854$	& $1E{-}05$	& $6E{-}09$	& $4E{-}10$	\\
    \midrule
    \texttt{rpd\_df\_1}	& $0.4240$	& $0.1895$	& $0.4543$	& $0.005$	& $8E{-}05$	& $3E{-}04$	 \\
    \texttt{rpd\_df\_2}	& $0.4200$	& $0.1972$	& $0.4727$	& $0.003$	& $1E{-}04$	& $9E{-}04$	 \\
    \texttt{rpd\_df\_3}	& $0.3880$	& $0.1757$	& $0.4304$	& $0.001$	& $2E{-}05$	& $8E{-}05$	 \\
    \texttt{rpd\_df\_4}	& $0.3360$	& $0.1458$	& $0.4000$	& $7E{-}05$	& $8E{-}07$	& $6E{-}06$	 \\
    \texttt{rpd\_df\_5}	& $0.2960$	& $0.1140$	& $0.3495$	& $9E{-}06$	& $1E{-}08$	& $1E{-}07$	\\
    \midrule
    \texttt{rpd\_tol\_1}	& $0.4200$	& $0.1872$	& $0.4469$	& $0.005$	& $6E{-}05$	& $2E{-}04$	 \\
    \texttt{rpd\_tol\_2}	& $0.3960$	& $0.1769$	& $0.4134$	& $0.002$	& $3E{-}05$	& $5E{-}05$	 \\
    \texttt{rpd\_tol\_3}	& $0.2040$	& $0.0987$	& $0.2365$	& $7E{-}08$	& $8E{-}09$	& $1E{-}10$	\\
    \texttt{rpd\_tol\_4}	& $0.0720$	& $0.0183$	& $0.0572$	& $1E{-}12$	& $5E{-}14$	& $3E{-}22$	\\
    \texttt{rpd\_tol\_5}	& $0.0200$	& $0.0007$	& $0.0048$	& $5E{-}16$	& $1E{-}15$	& $3E{-}27$	 \\
    \midrule
    \texttt{rpd\_C\_1}	& $0.2600$	& $0.1228$	& $0.2786$	& $5E{-}06$	& $3E{-}07$	& $2E{-}08$	 \\
    \texttt{rpd\_C\_2}	& $0.2600$	& $0.1216$	& $0.2790$	& $5E{-}06$	& $2E{-}07$	& $2E{-}08$	\\
    \texttt{rpd\_C\_3}	& $0.2360$	& $0.0969$	& $0.2507$	& $8E{-}07$	& $7E{-}09$	& $5E{-}10$	\\
    \texttt{rpd\_C\_4}	& $0.3600$	& $0.1609$	& $0.4095$	& $3E{-}04$	& $4E{-}06$	& $1E{-}05$	\\
    \texttt{rpd\_C\_5}	& $0.3520$	& $0.1565$	& $0.4026$	& $2E{-}04$	& $2E{-}06$	& $8E{-}06$	 \\
    \bottomrule
\end{tabular}%
}
\end{table}

\subsection{Correlation Analysis}
\label{subsec:correlation_analysis}

\subsubsection*{Replicability}

Note that for some measures, namely Kendall's $\tau$, \ac{RBO}, $p$-value, the higher the score the better the replicated run, conversely for \ac{RMSE} and Delta \ac{ARP} (absolute difference in \ac{ARP}), the lower the score the better the replicated run. Thus, before computing the correlation among measures, we ensure that all the measure scores are consistent with respect to each other. Practically we consider the opposite of $\tau$, \ac{RBO} and $p$-values, and for \ac{ER} we consider $|1 - ER|$, since the closer its score to $1$, the better the replicability performance.

Table~\ref{tab:replicability_correlation} reports Kendall's $\tau$ correlation for replicability measures on the set of runs replicating \texttt{WCrobust04} (upper triangle, white background) and \texttt{WCrobust0405} (lower triangle, turquoise background). The correlation between \ac{ARP} and $\tau$ is low, below $0.29$, and higher for \ac{RBO} $0.70$. This validates the findings from Section~\ref{subsec:performance_analysis}, showing that Kendall's $\tau$ assumes a totally different perspective when evaluating replicability runs. Between $\tau$ and \ac{RBO}, \ac{RBO} correlates more with \ac{ARP} than 
$\tau$, especially with respect to \ac{AP} and \ac{nDCG}. Also, $\tau$ and \ac{RBO} are low correlated with respect to each other. This is due to \ac{RBO} being top-heavy, as \ac{AP} and \ac{nDCG}, while Kendall's $\tau$ considers each rank position as equally important.

The correlation among \ac{ARP} and \ac{RMSE} is higher, especially when the same measure is considered by both \ac{ARP} and \ac{RMSE}. Nevertheless, the correlation is always lower than $0.86$, showing that it is different to compare the overall average or the performance score topic by topic, as also shown by P@10 in Table~\ref{tab:replicability_04}. Furthermore, the correlation between \ac{RMSE} instantiated with \ac{AP} and \ac{nDCG} is high, above $0.9$, this is due to \ac{AP} and \ac{nDCG} being highly correlated, as also shown by the correlation between \ac{ARP} with \ac{AP} and \ac{nDCG} (above $0.90$) and between $p$-values with \ac{AP} and \ac{nDCG} (above $0.91$).

When using the same performance measure, \ac{ARP} and $p$-values approaches are highly correlated, even if from Table~\ref{tab:replicability_04} 
several runs have small $p$-values and are statistically different. As mentioned in Section~\ref{subsec:performance_analysis}, the numerator of the $t$-stat is Delta \ac{ARP}, and likely due to low variance, Delta \ac{ARP} and $p$-values are tightly related. 

As explained in Section~\ref{subsec:replicability}, \ac{ER} takes a different perspective when evaluating replicability runs. This is corroborated by correlation results, which show that this measure has low correlation with \ac{ARP} and any other evaluation approach. Indeed, replicating the overall improvement over a baseline, does not mean that there is perfect replication on each topic. Moreover, even the correlation among \ac{ER} instantiated with different measures is low, which means that a mean improvement over the baseline in terms of \ac{AP} does not necessarily correspond to a similar mean improvement for \ac{nDCG}.

\begin{table*}[tb]
\caption{Replicability: correlation among different measures for runs replicating \texttt{WCrobust04} (white background); and runs replicating \texttt{WCrobust0405} (turquoise background).}
\label{tab:replicability_correlation}
\resizebox{\textwidth}{!}{
\begin{tabular}{@{}l|ccc|cc|ccc|ccc|ccc@{}}
\toprule
 & \multicolumn{3}{c|}{Delta \ac{ARP}} & \multicolumn{2}{c|}{Correlation} & \multicolumn{3}{c|}{\ac{RMSE}} & \multicolumn{3}{c|}{$p$-value} & \multicolumn{3}{c}{\ac{ER}} \\
 & P@10   & \ac{AP}    & \ac{nDCG}   & $\tau$   & \ac{RBO}  & P@10 & \ac{AP}    & \ac{nDCG}  & P@10  & \ac{AP} & \ac{nDCG}   & P@10 & \ac{AP}    & \ac{nDCG} \\
\midrule
$\Delta$arp\_P@10	& \cellcolor{lg}-	& $0.4175$	& $0.3979$	& $0.2456$	& $0.3684$	& $0.3419$	& $0.4552$	& $0.4290$	& $0.9156$	& $0.3668$	& $0.3700$	& $0.2348$	& $0.1752$	& $0.0884$\\
$\Delta$arp\_\ac{AP}	& \cellcolor{pt}$0.4535$	& \cellcolor{lg}-	& $0.9118$	& $0.2718$	& $0.7045$	& $0.5209$	& $0.8514$	& $0.8090$	& $0.3855$	& $0.8841$	& $0.8596$	& $0.2145$	& $0.3012$	& $0.3731$\\
$\Delta$arp\_\ac{nDCG}	& \cellcolor{pt}$0.4716$	& \cellcolor{pt}$0.9363$	& \cellcolor{lg}-	& $0.2882$	& $0.6555$	& $0.5339$	& $0.8580$	& $0.8547$	& $0.3463$	& $0.8318$	& $0.8302$	& $0.2374$	& $0.3208$	& $0.4318$\\
\midrule
$\tau$	& \cellcolor{pt}$0.2620$	& \cellcolor{pt}$0.2865$	& \cellcolor{pt}$0.2620$	& \cellcolor{lg}-	& $0.2180$	& $0.2788$	& $0.2702$	& $0.2898$	& $0.2434$	& $0.2376$	& $0.2457$	& $0.1834$	& $0.2718$	& $0.2098$\\
\ac{RBO}	& \cellcolor{pt}$0.3946$	& \cellcolor{pt}$0.6637$	& \cellcolor{pt}$0.6457$	& \cellcolor{pt}$0.3584$	& \cellcolor{lg}-	& $0.6026$	& $0.7616$	& $0.6898$	& $0.3201$	& $0.6376$	& $0.6490$	& $0.3307$	& $0.2049$	& $0.3029$\\
\midrule
\ac{RMSE}\_P@10	& \cellcolor{pt}$0.5420$	& \cellcolor{pt}$0.6713$	& \cellcolor{pt}$0.7089$	& \cellcolor{pt}$0.3213$	& \cellcolor{pt}$0.7433$	& \cellcolor{lg}-	& $0.6239$	& $0.5944$	& $0.2544$	& $0.4080$	& $0.4129$	& $0.3452$	& $0.2706$	& $0.3753$\\
\ac{RMSE}\_\ac{AP}	& \cellcolor{pt}$0.5076$	& \cellcolor{pt}$0.7747$	& \cellcolor{pt}$0.8188$	& \cellcolor{pt}$0.3224$	& \cellcolor{pt}$0.7910$	& \cellcolor{pt}$0.8136$	& \cellcolor{lg}-	& $0.8988$	& $0.4034$	& $0.7355$	& $0.7273$	& $0.2734$	& $0.3453$	& $0.4171$\\
\ac{RMSE}\_\ac{nDCG}	& \cellcolor{pt}$0.4666$	& \cellcolor{pt}$0.7616$	& \cellcolor{pt}$0.8188$	& \cellcolor{pt}$0.3094$	& \cellcolor{pt}$0.7682$	& \cellcolor{pt}$0.8054$	& \cellcolor{pt}$0.9184$	& \cellcolor{lg}-	& $0.3806$	& $0.7127$	& $0.6849$	& $0.2767$	& $0.3649$	& $0.4498$\\
\midrule
p\_value\_P@10	& \cellcolor{pt}$0.8393$	& \cellcolor{pt}$0.3694$	& \cellcolor{pt}$0.3645$	& \cellcolor{pt}$0.2566$	& \cellcolor{pt}$0.2877$	& \cellcolor{pt}$0.3790$	& \cellcolor{pt}$0.3743$	& \cellcolor{pt}$0.3400$	& \cellcolor{lg}-	& $0.3740$	& $0.3593$	& $0.2129$	& $0.1486$	& $0.0327$\\
p\_value\_\ac{AP}	& \cellcolor{pt}$0.3913$	& \cellcolor{pt}$0.8498$	& \cellcolor{pt}$0.7927$	& \cellcolor{pt}$0.2506$	& \cellcolor{pt}$0.5657$	& \cellcolor{pt}$0.5470$	& \cellcolor{pt}$0.6245$	& \cellcolor{pt}$0.6180$	& \cellcolor{pt}$0.3564$	& \cellcolor{lg}-	& $0.9135$	& $0.1736$	& $0.2343$	& $0.2898$\\
p\_value\_\ac{nDCG}	& \cellcolor{pt}$0.3848$	& \cellcolor{pt}$0.8416$	& \cellcolor{pt}$0.7845$	& \cellcolor{pt}$0.2424$	& \cellcolor{pt}$0.5543$	& \cellcolor{pt}$0.5356$	& \cellcolor{pt}$0.6196$	& \cellcolor{pt}$0.6033$	& \cellcolor{pt}$0.3384$	& \cellcolor{pt}$0.9069$	& \cellcolor{lg}-	& $0.2178$	& $0.2163$	& $0.3110$\\
\midrule
\ac{ER}\_P@10	& \cellcolor{pt}$0.0739$	& \cellcolor{pt}$0.2652$	& \cellcolor{pt}$0.2767$	& \cellcolor{pt}$0.2227$	& \cellcolor{pt}$0.3537$	& \cellcolor{pt}$0.3108$	& \cellcolor{pt}$0.3193$	& \cellcolor{pt}$0.3144$	& \cellcolor{pt}$0.0459$	& \cellcolor{pt}$0.1817$	& \cellcolor{pt}$0.1867$	& \cellcolor{lg}-	& $0.2833$	& $0.1736$\\
\ac{ER}\_\ac{AP}	& \cellcolor{pt}$0.3013$	& \cellcolor{pt}$0.2963$	& \cellcolor{pt}$0.3078$	& \cellcolor{pt}$0.1673$	& \cellcolor{pt}$0.2343$	& \cellcolor{pt}$0.3312$	& \cellcolor{pt}$0.3551$	& \cellcolor{pt}$0.3420$	& \cellcolor{pt}$0.2599$	& \cellcolor{pt}$0.1886$	& \cellcolor{pt}$0.1706$	& \cellcolor{pt}$0.2833$	& \cellcolor{lg}-	& $0.3992$\\
\ac{ER}\_\ac{nDCG}	& \cellcolor{pt}$0.2718$	& \cellcolor{pt}$0.2767$	& \cellcolor{pt}$0.3143$	& \cellcolor{pt}$0.1216$	& \cellcolor{pt}$0.2669$	& \cellcolor{pt}$0.3377$	& \cellcolor{pt}$0.3747$	& \cellcolor{pt}$0.3551$	& \cellcolor{pt}$0.1553$	& \cellcolor{pt}$0.1494$	& \cellcolor{pt}$0.1706$	& \cellcolor{pt}$0.1736$	& \cellcolor{pt}$0.3992$	& \cellcolor{lg}- \\
\bottomrule
\end{tabular}
}
\end{table*}

\subsubsection*{Reproducibility}

\begin{table}[tb]
\caption{Reproducibility: correlation among different measures for runs reproducing \texttt{WCrobust04} (white background); and runs reproducing \texttt{WCrobust0405} (turquoise background).}
\label{tab:reproducibility_correlation}
\resizebox{0.49\textwidth}{!}{
\begin{tabular}{@{}l|ccc|ccc@{}}
\toprule
 & \multicolumn{3}{c|}{$p$-value} & \multicolumn{3}{c}{\ac{ER}} \\
 & P@10   & \ac{AP}   & \ac{nDCG}   & P@10   & \ac{AP}   & \ac{nDCG} \\
\midrule
p\_value\_P@10	& \cellcolor{lg}-	& $0.8545$	& $0.8446$	& $-0.2050$	& $-0.1153$	& $0.0025$\\
p\_value\_\ac{AP}	& \cellcolor{pt}$0.8168$	& \cellcolor{lg}-	& $0.8694$	& $-0.1743$	& $-0.1151$	& $-0.0335$\\
p\_value\_\ac{nDCG}	& \cellcolor{pt}$0.8054$	& \cellcolor{pt}$0.9216$	& \cellcolor{lg}-	& $-0.2350$	& $-0.2033$	& $-0.0857$\\
\midrule
\ac{ER}\_P@10	& \cellcolor{pt}$0.0939$	& \cellcolor{pt}$0.0674$	& \cellcolor{pt}$0.0756$	& \cellcolor{lg}-	& $0.5651$	& $0.3091$\\
\ac{ER}\_\ac{AP} &	 \cellcolor{pt}$0.2232$	& \cellcolor{pt}$0.2082$	& \cellcolor{pt}$0.2473$	& \cellcolor{pt}$0.5886$	& \cellcolor{lg}-	& $0.5298$\\
\ac{ER}\_\ac{nDCG}	& \cellcolor{pt}$0.1006$	& \cellcolor{pt}$0.1167$	& \cellcolor{pt}$0.1559$	& \cellcolor{pt}$0.2220$	& \cellcolor{pt}$0.4318$	& \cellcolor{lg}- \\
\bottomrule
\end{tabular}
}
\end{table}

For reproducibility we can not compare against \ac{ARP}: since the original and reproduced runs are defined on different collections, it is meaningless to contrast average scores. Table~\ref{tab:reproducibility_correlation} reports the correlation among reproducibility runs for \texttt{WCrobust04} (upper triangle, white background) and for \texttt{WC\-ro\-bust\-04\-05} (lower triangle, turquoise background). Again, before computing the correlation among different measures, we ensured that the meaning of their scores is consistent across measures, i.e.~the lower the score the better the reproduced results.

The correlation results for reproducibility show once more that \ac{ER} is low correlated to $p$-values approaches, thus these methods are taking two different evaluation perspectives. Furthermore, \ac{ER} has low correlation with itself when instantiated with different performance measures: even for reproducibility, two different performance measures do not exhibit an average improvement over baseline runs in a similar way.

Finally, all $p$-values approaches are fairly correlated with respect to each other, even stronger than in the replicability case of Table~\ref{tab:replicability_correlation}. This is surprising, if we consider that all the reproducibility runs are statistically significantly different, as shown in Table~\ref{tab:reproducibility_p_value}. However, it represents a further signal that the unpaired $t$-test is not able to recognise successfully reproduced runs, when the new collection and the original collection are too different, independently of the effectiveness measure.

\section{Conclusions and Future Work}
\label{sec:conclusions}

We faced the core issue of investigating measures to determine to what extent a system-oriented \ac{IR} experiment has been replicated or reproduced. To this end, we analysed and compared several measures at different levels of granularity and we developed the first reproducibility-oriented dataset. Due to the lack of a reproducibility-oriented dataset, these measures have never been validated so far. 

We found that replicability measures behave as expected and consistently; in particular, \ac{RBO} provides more meaningfull comparisons than Kendall's $\tau$; \ac{RMSE} properly indicates whether we obtained a similar level of performance; finally, both \ac{ER}/\ac{DeltaRI} and the paired t-test successfully determine whether the same effects are replicated. On the other hand, quantifying reproducibility is more challenging and, while \ac{ER}/\ac{DeltaRI} are still able to provide sensible insights, the unpaired t-test seems to be too sensitive to the differences among the experimental collections.

As a suggestion to improve our community practices, it is important to always provide not only the source code but also the actual run, as to enable  precise checking for replicability; luckily, this is already happening when we operate within evaluation campaigns which gather and make available runs by their participants.

In future work,
we will explore more advanced statistical methods to quantify reproducibility in a reliable way. Moreover, we will investigate how replicability and reproducibility are related to user experience. For example, a perfectly replicated run in terms of \ac{RMSE}, but with low \ac{RBO}, presents different documents to a user and this might greatly affect her/his experience. Therefore, we need to better understand which replicability/reproducibility level is needed to not impact (too much) on the user experience.

\vspace{0.5em}
\small
\noindent \textbf{Acknowledgments}. This paper is partially supported by AMAOS (Advanced Machine Learning for Automatic Omni-Channel Support), funded by Innovationsfonden, Denmark, and by DFG (German Research Foundation, project no. 407518790).

\acrodef{3G}[3G]{Third Generation Mobile System}
\acrodef{5S}[5S]{Streams, Structures, Spaces, Scenarios, Societies}
\acrodef{AAAI}[AAAI]{Association for the Advancement of Artificial Intelligence}
\acrodef{AAL}[AAL]{Annotation Abstraction Layer}
\acrodef{AAM}[AAM]{Automatic Annotation Manager}
\acrodef{ACLIA}[ACLIA]{Advanced Cross-Lingual Information Access}
\acrodef{ACM}[ACM]{Association for Computing Machinery}
\acrodef{ADSL}[ADSL]{Asymmetric Digital Subscriber Line}
\acrodef{ADUI}[ADUI]{ADministrator User Interface}
\acrodef{AIP}[AIP]{Archival Information Package}
\acrodef{AJAX}[AJAX]{Asynchronous JavaScript Technology and \acs{XML}}
\acrodef{ALU}[ALU]{Aritmetic-Logic Unit}
\acrodef{AMUSID}[AMUSID]{Adaptive MUSeological IDentity-service}
\acrodef{ANOVA}[ANOVA]{ANalysis Of VAriance}
\acrodef{ANSI}[ANSI]{American National Standards Institute}
\acrodef{AP}[AP]{Average Precision}
\acrodef{APC}[APC]{AP Correlation}
\acrodef{API}[API]{Application Program Interface}
\acrodef{AR}[AR]{Address Register}
\acrodef{AS}[AS]{Annotation Service}
\acrodef{ASAP}[ASAP]{Adaptable Software Architecture Performance}
\acrodef{ASI}[ASI]{Annotation Service Integrator}
\acrodef{ASL}[ASL]{Achieved Significance Level}
\acrodef{ASM}[ASM]{Annotation Storing Manager}
\acrodef{ASR}[ASR]{Automatic Speech Recognition}
\acrodef{ASUI}[ASUI]{ASsessor User Interface}
\acrodef{ATIM}[ATIM]{Annotation Textual Indexing Manager}
\acrodef{AUC}[AUC]{Area Under the ROC Curve}
\acrodef{AUI}[AUI]{Administrative User Interface}
\acrodef{ARP}[ARP]{Average Retrieval Performance}
\acrodef{AWARE}[AWARE]{Assessor-driven Weighted Averages for Retrieval Evaluation}
\acrodef{BANKS-I}[BANKS-I]{Browsing ANd Keyword Searching I}
\acrodef{BANKS-II}[BANKS-II]{Browsing ANd Keyword Searching II}
\acrodef{bpref}[bpref]{Binary Preference}
\acrodef{BNF}[BNF]{Backus and Naur Form}
\acrodef{BRICKS}[BRICKS]{Building Resources for Integrated Cultural Knowledge Services}
\acrodef{CAN}[CAN]{Content Addressable Netword}
\acrodef{CAS}[CAS]{Content-And-Structure}
\acrodef{CBSD}[CBSD]{Component-Based Software Developlement}
\acrodef{CBSE}[CBSE]{Component-Based Software Engineering}
\acrodef{CB-SPE}[CB-SPE]{Component-Based \acs{SPE}}
\acrodef{CD}[CD]{Collaboration Diagram}
\acrodef{CD}[CD]{Compact Disk}
\acrodef{CDF}[CDF]{Cumulative Density Function}
\acrodef{CENL}[CENL]{Conference of European National Librarians}
\acrodef{CIDOC CRM}[CIDOC CRM]{CIDOC Conceptual Reference Model}
\acrodef{CIR}[CIR]{Current Instruction Register}
\acrodef{CIRCO}[CIRCO]{Coordinated Information Retrieval Components Orchestration}
\acrodef{CG}[CG]{Cumulated Gain}
\acrodef{CL}[CL]{Curriculum Learning}
\acrodef{CL-ESA}[CL-ESA]{Cross-Lingual Explicit Semantic Analysis}
\acrodef{CLAIRE}[CLAIRE]{Combinatorial visuaL Analytics system for Information Retrieval Evaluation}
\acrodef{CLEF1}[CLEF]{Cross-Language Evaluation Forum}
\acrodef{CLEF}[CLEF]{Conference and Labs of the Evaluation Forum}
\acrodef{CLIR}[CLIR]{Cross Language Information Retrieval}
\acrodef{CM}[CM]{Continuation Methods}
\acrodef{CMS}[CMS]{Content Management System}
\acrodef{CMT}[CMT]{Campaign Management Tool}
\acrodef{CNR}[CNR]{Italian National Council of Research}
\acrodef{CO}[CO]{Content-Only}
\acrodef{COD}[COD]{Code On Demand}
\acrodef{CODATA}[CODATA]{Committee on Data for Science and Technology}
\acrodef{COLLATE}[COLLATE]{Collaboratory for Annotation Indexing and Retrieval of Digitized Historical Archive Material}
\acrodef{CP}[CP]{Characteristic Pattern}
\acrodef{CPE}[CPE]{Control Processor Element}
\acrodef{CPU}[CPU]{Central Processing Unit}
\acrodef{CQL}[CQL]{Contextual Query Language}
\acrodef{CRP}[CRP]{Cumulated Relative Position}
\acrodef{CRUD}[CRUD]{Create--Read--Update--Delete}
\acrodef{CS}[CS]{Characteristic Structure}
\acrodef{CSM}[CSM]{Campaign Storing Manager}
\acrodef{CSS}[CSS]{Cascading Style Sheets}
\acrodef{CTR}[CTR]{Click-Through Rate}
\acrodef{CU}[CU]{Control Unit}
\acrodef{CUI}[CUI]{Client User Interface}
\acrodef{CV}[CV]{Cross-Validation}
\acrodef{DAFFODIL}[DAFFODIL]{Distributed Agents for User-Friendly Access of Digital Libraries}
\acrodef{DAO}[DAO]{Data Access Object}
\acrodef{DARE}[DARE]{Drawing Adequate REpresentations}
\acrodef{DARPA}[DARPA]{Defense Advanced Research Projects Agency}
\acrodef{DAS}[DAS]{Distributed Annotation System}
\acrodef{DB}[DB]{DataBase}
\acrodef{DBMS}[DBMS]{DataBase Management System}
\acrodef{DC}[DC]{Dublin Core}
\acrodef{DCG}[DCG]{Discounted Cumulated Gain}
\acrodef{DCMI}[DCMI]{Dublin Core Metadata Initiative}
\acrodef{DCV}[DCV]{Document Cut--off Value}
\acrodef{DD}[DD]{Deployment Diagram}
\acrodef{DDC}[DDC]{Dewey Decimal Classification}
\acrodef{DDS}[DDS]{Direct Data Structure}
\acrodef{DF}[DF]{Degrees of Freedom}
\acrodef{DFI}[DFI]{Divergence From Independence}
\acrodef{DFR}[DFR]{Divergence From Randomness}
\acrodef{DHT}[DHT]{Distributed Hash Table}
\acrodef{DI}[DI]{Digital Image}
\acrodef{DIKW}[DIKW]{Data, Information, Knowledge, Wisdom}
\acrodef{DIL}[DIL]{\acs{DIRECT} Integration Layer}
\acrodef{DiLAS}[DiLAS]{Digital Library Annotation Service}
\acrodef{DIRECT}[DIRECT]{Distributed Information Retrieval Evaluation Campaign Tool}
\acrodef{DKMS}[DKMS]{Data and Knowledge Management System}
\acrodef{DL}[DL]{Digital Library}
\acrodefplural{DL}[DL]{Digital Libraries}
\acrodef{DLMS}[DLMS]{Digital Library Management System}
\acrodef{DLOG}[DL]{Description Logics}
\acrodef{DLS}[DLS]{Digital Library System}
\acrodef{DLSS}[DLSS]{Digital Library Service System}
\acrodef{DM}[DM]{Data Mining}
\acrodef{DO}[DO]{Digital Object}
\acrodef{DOI}[DOI]{Digital Object Identifier}
\acrodef{DOM}[DOM]{Document Object Model}
\acrodef{DoMDL}[DoMDL]{Document Model for Digital Libraries}
\acrodef{DP}[DP]{Discriminative Power}
\acrodef{DPBF}[DPBF]{Dynamic Programming Best-First}
\acrodef{DR}[DR]{Data Register}
\acrodef{DRIVER}[DRIVER]{Digital Repository Infrastructure Vision for European Research}
\acrodef{DTD}[DTD]{Document Type Definition}
\acrodef{DVD}[DVD]{Digital Versatile Disk}
\acrodef{EAC-CPF}[EAC-CPF]{Encoded Archival Context for Corporate Bodies, Persons, and Families}
\acrodef{EAD}[EAD]{Encoded Archival Description}
\acrodef{EAN}[EAN]{International Article Number}
\acrodef{EBU}[EBU]{Expected Browsing Utility}
\acrodef{ECD}[ECD]{Enhanced Contenty Delivery}
\acrodef{ECDL}[ECDL]{European Conference on Research and Advanced Technology for Digital Libraries}
\acrodef{EDM}[EDM]{Europeana Data Model}
\acrodef{EG}[EG]{Execution Graph}
\acrodef{ELDA}[ELDA]{Evaluation and Language resources Distribution Agency}
\acrodef{ELRA}[ELRA]{European Language Resources Association}
\acrodef{EM}[EM]{Expectation Maximization}
\acrodef{EMMA}[EMMA]{Extensible MultiModal Annotation}
\acrodef{EPROM}[EPROM]{Erasable Programmable \acs{ROM}}
\acrodef{EQNM}[EQNM]{Extended Queueing Network Model}
\acrodef{ERR}[ERR]{Expected Reciprocal Rank}
\acrodef{ESA}[ESA]{Explicit Semantic Analysis}
\acrodef{ESL}[ESL]{Expected Search Length}
\acrodef{ETL}[ETL]{Extract-Transform-Load}
\acrodef{FAST}[FAST]{Flexible Annotation Service Tool}
\acrodef{FDR}[FDR]{False Discovery Rate}
\acrodef{FIFO}[FIFO]{First-In / First-Out}
\acrodef{FIRE}[FIRE]{Forum for Information Retrieval Evaluation}
\acrodef{FN}[FN]{False Negative}
\acrodef{FNR}[FNR]{False Negative Rate}
\acrodef{FOAF}[FOAF]{Friend of a Friend}
\acrodef{FORESEE}[FORESEE]{FOod REcommentation sErvER}
\acrodef{FP}[FP]{False Positive}
\acrodef{FPR}[FPR]{False Positive Rate}
\acrodef{FWER}[FWER]{Family-wise Error Rate}
\acrodef{GIF}[GIF]{Graphics Interchange Format}
\acrodef{GIR}[GIR]{Geografic Information Retrieval}
\acrodef{GAP}[GAP]{Graded Average Precision}
\acrodef{GLM}[GLM]{General Linear Model}
\acrodef{GLMM}[GLMM]{General Linear Mixed Model}
\acrodef{GMAP}[GMAP]{Geometric Mean Average Precision}
\acrodef{GoP}[GoP]{Grid of Points}
\acrodef{GPRS}[GPRS]{General Packet Radio Service}
\acrodef{gP}[gP]{Generalized Precision}
\acrodef{gR}[gR]{Generalized Recall}
\acrodef{gRBP}[gRBP]{Graded Rank-Biased Precision}
\acrodef{GT}[GT]{Generalizability Theory}
\acrodef{GTIN}[GTIN]{Global Trade Item Number}
\acrodef{GUI}[GUI]{Graphical User Interface}
\acrodef{GW}[GW]{Gateway}
\acrodef{HCI}[HCI]{Human Computer Interaction}
\acrodef{HDS}[HDS]{Hybrid Data Structure}
\acrodef{HIR}[HIR]{Hypertext Information Retrieval}
\acrodef{HIT}[HIT]{Human Intelligent Task}
\acrodef{HITS}[HITS]{Hyperlink-Induced Topic Search}
\acrodef{HMM}[HMM]{Hidden Markov Model}
\acrodef{HTML}[HTML]{HyperText Markup Language}
\acrodef{HTTP}[HTTP]{HyperText Transfer Protocol}
\acrodef{HSD}[HSD]{Honestly Significant Difference}
\acrodef{ICA}[ICA]{International Council on Archives}
\acrodef{ICSU}[ICSU]{International Council for Science}
\acrodef{IDF}[IDF]{Inverse Document Frequency}
\acrodef{IDS}[IDS]{Inverse Data Structure}
\acrodef{IEEE}[IEEE]{Institute of Electrical and Electronics Engineers}
\acrodef{IEI}[IEI]{Istituto della Enciclopedia Italiana fondata da Giovanni Treccani}
\acrodef{IETF}[IETF]{Internet Engineering Task Force}
\acrodef{IIR}[IIR]{Interactive Information Retrieval}
\acrodef{IMS}[IMS]{Information Management System}
\acrodef{IMSPD}[IMS]{Information Management Systems Research Group}
\acrodef{indAP}[indAP]{Induced Average Precision}
\acrodef{infAP}[infAP]{Inferred Average Precision}
\acrodef{INEX}[INEX]{INitiative for the Evaluation of \acs{XML} Retrieval}
\acrodef{INS-M}[INS-M]{Inverse Set Data Model}
\acrodef{INTR}[INTR]{Interrupt Register}
\acrodef{IP}[IP]{Internet Protocol}
\acrodef{IPSA}[IPSA]{Imaginum Patavinae Scientiae Archivum}
\acrodef{IR}[IR]{Information Retrieval}
\acrodef{IRON}[IRON]{Information Retrieval ON}
\acrodef{IRON2}[IRON$^2$]{Information Retrieval On aNNotations}
\acrodef{IRON-SAT}[IRON-SAT]{\acs{IRON} - Statistical Analysis Tool}
\acrodef{IRS}[IRS]{Information Retrieval System}
\acrodef{ISAD(G)}[ISAD(G)]{International Standard for Archival Description (General)}
\acrodef{ISBN}[ISBN]{International Standard Book Number}
\acrodef{ISIS}[ISIS]{Interactive SImilarity Search}
\acrodef{ISJ}[ISJ]{Interactive Searching and Judging}
\acrodef{ISO}[ISO]{International Organization for Standardization}
\acrodef{ITU}[ITU]{International Telecommunication Union }
\acrodef{ITU-T}[ITU-T]{Telecommunication Standardization Sector of \acs{ITU}}
\acrodef{IV}[IV]{Information Visualization}
\acrodef{JAN}[JAN]{Japanese Article Number}
\acrodef{JDBC}[JDBC]{Java DataBase Connectivity}
\acrodef{JMB}[JMB]{Java--Matlab Bridge}
\acrodef{JPEG}[JPEG]{Joint Photographic Experts Group}
\acrodef{JSON}[JSON]{JavaScript Object Notation}
\acrodef{JSP}[JSP]{Java Server Pages}
\acrodef{JTE}[JTE]{Java-Treceval Engine}
\acrodef{KDE}[KDE]{Kernel Density Estimation}
\acrodef{KLD}[KLD]{Kullback-Leibler Divergence}
\acrodef{KLAPER}[KLAPER]{Kernel LAnguage for PErformance and Reliability analysis}
\acrodef{LAM}[LAM]{Libraries, Archives, and Museums}
\acrodef{LAM2}[LAM]{Logistic Average Misclassification}
\acrodef{LAN}[LAN]{Local Area Network}
\acrodef{LD}[LD]{Linked Data}
\acrodef{LEAF}[LEAF]{Linking and Exploring Authority Files}
\acrodef{LIDO}[LIDO]{Lightweight Information Describing Objects}
\acrodef{LIFO}[LIFO]{Last-In / First-Out}
\acrodef{LM}[LM]{Language Model}
\acrodef{LMT}[LMT]{Log Management Tool}
\acrodef{LOD}[LOD]{Linked Open Data}
\acrodef{LODE}[LODE]{Linking Open Descriptions of Events}
\acrodef{LpO}[LpO]{Leave-$p$-Out}
\acrodef{LRM}[LRM]{Local Relational Model}
\acrodef{LRU}[LRU]{Last Recently Used}
\acrodef{LS}[LS]{Lexical Signature}
\acrodef{LSM}[LSM]{Log Storing Manager}
\acrodef{LtR}[LtR]{Learning to Rank}
\acrodef{LUG}[LUG]{Lexical Unit Generator}
\acrodef{MA}[MA]{Mobile Agent}
\acrodef{MA}[MA]{Moving Average}
\acrodef{MACS}[MACS]{Multilingual ACcess to Subjects}
\acrodef{MADCOW}[MADCOW]{Multimedia Annotation of Digital Content Over the Web}
\acrodef{MAD}[MAD]{Mean Assessed Documents}
\acrodef{MADP}[MADP]{Mean Assessed Documents Precision}
\acrodef{MADS}[MADS]{Metadata Authority Description Standard}
\acrodef{MAP}[MAP]{Mean Average Precision}
\acrodef{MARC}[MARC]{Machine Readable Cataloging}
\acrodef{MATTERS}[MATTERS]{MATlab Toolkit for Evaluation of information Retrieval Systems}
\acrodef{MDA}[MDA]{Model Driven Architecture}
\acrodef{MDD}[MDD]{Model-Driven Development}
\acrodef{METS}[METS]{Metadata Encoding and Transmission Standard}
\acrodef{MIDI}[MIDI]{Musical Instrument Digital Interface}
\acrodef{MIME}[MIME]{Multipurpose Internet Mail Extensions}
\acrodef{ML}[ML]{Machine Learning}
\acrodef{MLE}[MLE]{Maximum Likelihood Estimation}
\acrodef{MLIA}[MLIA]{MultiLingual Information Access}
\acrodef{MM}[MM]{Machinery Model}
\acrodef{MMU}[MMU]{Memory Management Unit}
\acrodef{MODS}[MODS]{Metadata Object Description Schema}
\acrodef{MOF}[MOF]{Meta-Object Facility}
\acrodef{MP}[MP]{Markov Precision}
\acrodef{MPEG}[MPEG]{Motion Picture Experts Group}
\acrodef{MRD}[MRD]{Machine Readable Dictionary}
\acrodef{MRF}[MRF]{Markov Random Field}
\acrodef{MRR}[MRR]{Mean Reciprocal Rank}
\acrodef{MS}[MS]{Mean Squares}
\acrodef{MSAC}[MSAC]{Multilingual Subject Access to Catalogues}
\acrodef{MSE}[MSE]{Mean Square Error}
\acrodef{MT}[MT]{Machine Translation}
\acrodef{MV}[MV]{Majority Vote}
\acrodef{MVC}[MVC]{Model-View-Controller}
\acrodef{NACSIS}[NACSIS]{NAtional Center for Science Information Systems}
\acrodef{NAP}[NAP]{Network processors Applications Profile}
\acrodef{NCP}[NCP]{Normalized Cumulative Precision}
\acrodef{nCG}[nCG]{Normalized Cumulated Gain}
\acrodef{nCRP}[nCRP]{Normalized Cumulated Relative Position}
\acrodef{nDCG}[nDCG]{Normalized Discounted Cumulated Gain}
\acrodef{nMCG}[nMCG]{Normalized Markov Cumulated Gain}
\acrodef{NESTOR}[NESTOR]{NEsted SeTs for Object hieRarchies}
\acrodef{NEXI}[NEXI]{Narrowed Extended XPath I}
\acrodef{NII}[NII]{National Institute of Informatics}
\acrodef{NISO}[NISO]{National Information Standards Organization}
\acrodef{NIST}[NIST]{National Institute of Standards and Technology}
\acrodef{NLP}[NLP]{Natural Language Processing}
\acrodef{NN}[NN]{Neural Network}
\acrodef{NP}[NP]{Network Processor}
\acrodef{NR}[NR]{Normalized Recall}
\acrodef{NS-M}[NS-M]{Nested Set Model}
\acrodef{NTCIR}[NTCIR]{NII Testbeds and Community for Information access Research}
\acrodef{OAI}[OAI]{Open Archives Initiative}
\acrodef{OAI-ORE}[OAI-ORE]{Open Archives Initiative Object Reuse and Exchange}
\acrodef{OAI-PMH}[OAI-PMH]{Open Archives Initiative Protocol for Metadata Harvesting}
\acrodef{OAIS}[OAIS]{Open Archival Information System}
\acrodef{OC}[OC]{Operation Code}
\acrodef{OCLC}[OCLC]{Online Computer Library Center}
\acrodef{OMG}[OMG]{Object Management Group}
\acrodef{OO}[OO]{Object Oriented}
\acrodef{OODB}[OODB]{Object-Oriented \acs{DB}}
\acrodef{OODBMS}[OODBMS]{Object-Oriented \acs{DBMS}}
\acrodef{OPAC}[OPAC]{Online Public Access Catalog}
\acrodef{OQL}[OQL]{Object Query Language}
\acrodef{ORP}[ORP]{Open Relevance Project}
\acrodef{OSIRIS}[OSIRIS]{Open Service Infrastructure for Reliable and Integrated process Support}
\acrodef{P}[P]{Precision}
\acrodef{P2P}[P2P]{Peer-To-Peer}
\acrodef{PA}[PA]{Performance Analysis}
\acrodef{PAMT}[PAMT]{Pool-Assessment Management Tool}
\acrodef{PASM}[PASM]{Pool-Assessment Storing Manager}
\acrodef{PC}[PC]{Program Counter}
\acrodef{PCP}[PCP]{Pre-Commercial Procurement}
\acrodef{PCR}[PCR]{Peripherical Command Register}
\acrodef{PDA}[PDA]{Personal Digital Assistant}
\acrodef{PDF}[PDF]{Probability Density Function}
\acrodef{PDR}[PDR]{Peripherical Data Register}
\acrodef{PIR}[PIR]{Personalized Information Retrieval}
\acrodef{POI}[POI]{\acs{PURL}-based Object Identifier}
\acrodef{PoS}[PoS]{Part of Speech}
\acrodef{PPE}[PPE]{Programmable Processing Engine}
\acrodef{PREFORMA}[PREFORMA]{PREservation FORMAts for culture information/e-archives}
\acrodef{PRIMAD}[PRIMAD]{Platform, Research goal, Implementation, Method, Actor, and Data}
\acrodef{PRIMAmob-UML}[PRIMAmob-UML]{mobile \acs{PRIMA-UML}}
\acrodef{PRIMA-UML}[PRIMA-UML]{PeRformance IncreMental vAlidation in \acs{UML}}
\acrodef{PROM}[PROM]{Programmable \acs{ROM}}
\acrodef{PROMISE}[PROMISE]{Participative Research labOratory  for Multimedia and Multilingual Information Systems Evaluation}
\acrodef{pSQL}[pSQL]{propagate \acs{SQL}}
\acrodef{PUI}[PUI]{Participant User Interface}
\acrodef{PURL}[PURL]{Persistent \acs{URL}}
\acrodef{QA}[QA]{Question Answering}
\acrodef{QE}[QE]{Query Expansion}
\acrodef{QoS-UML}[QoS-UML]{\acs{UML} Profile for QoS and Fault Tolerance}
\acrodef{QPA}[QPA]{Query Performance Analyzer}
\acrodef{QPP}[QPP]{Query Performance Prediction}
\acrodef{R}[R]{Recall}
\acrodef{RAM}[RAM]{Random Access Memory}
\acrodef{RAMM}[RAM]{Random Access Machine}
\acrodef{RBO}[RBO]{Rank-Biased Overlap}
\acrodef{RBP}[RBP]{Rank-Biased Precision}
\acrodef{RBTO}[RBTO]{Rank-Based Total Order}
\acrodef{RDBMS}[RDBMS]{Relational \acs{DBMS}}
\acrodef{RDF}[RDF]{Resource Description Framework}
\acrodef{REST}[REST]{REpresentational State Transfer}
\acrodef{REV}[REV]{Remote Evaluation}
\acrodef{RF}[RF]{Relevance Feedback}
\acrodef{RFC}[RFC]{Request for Comments}
\acrodef{RIA}[RIA]{Reliable Information Access}
\acrodef{RMSE}[RMSE]{Root Mean Square Error}
\acrodef{RMT}[RMT]{Run Management Tool}
\acrodef{ROM}[ROM]{Read Only Memory}
\acrodef{ROMIP}[ROMIP]{Russian Information Retrieval Evaluation Seminar}
\acrodef{RoMP}[RoMP]{Rankings of Measure Pairs}
\acrodef{RoS}[RoS]{Rankings of Systems}
\acrodef{RP}[RP]{Relative Position}
\acrodef{RR}[RR]{Reciprocal Rank}
\acrodef{RSM}[RSM]{Run Storing Manager}
\acrodef{RST}[RST]{Rhetorical Structure Theory}
\acrodef{RSV}[RSV]{Retrieval Status Value}
\acrodef{RT-UML}[RT-UML]{\acs{UML} Profile for Schedulability, Performance and Time}
\acrodef{SA}[SA]{Software Architecture}
\acrodef{SAL}[SAL]{Storing Abstraction Layer}
\acrodef{SAMT}[SAMT]{Statistical Analysis Management Tool}
\acrodef{SAN}[SAN]{Sistema Archivistico Nazionale}
\acrodef{SASM}[SASM]{Statistical Analysis Storing Manager}
\acrodef{SBTO}[SBTO]{Set-Based Total Order}
\acrodef{SD}[SD]{Sequence Diagram}
\acrodef{SE}[SE]{Search Engine}
\acrodef{SEBD}[SEBD]{Convegno Nazionale su Sistemi Evoluti per Basi di Dati}
\acrodef{SEM}[SEM]{Standard Error of the Mean}
\acrodef{SERP}[SERP]{Search Engine Result Page}
\acrodef{SFT}[SFT]{Satisfaction--Frustration--Total}
\acrodef{SIL}[SIL]{Service Integration Layer}
\acrodef{SIP}[SIP]{Submission Information Package}
\acrodef{SKOS}[SKOS]{Simple Knowledge Organization System}
\acrodef{SM}[SM]{Software Model}
\acrodef{SME}[SME]{Statistics--Metrics-Experiments}
\acrodef{SMART}[SMART]{System for the Mechanical Analysis and Retrieval of Text}
\acrodef{SoA}[SoA]{Service-oriented Architectures}
\acrodef{SOA}[SOA]{Strength of Association}
\acrodef{SOAP}[SOAP]{Simple Object Access Protocol}
\acrodef{SOM}[SOM]{Self-Organizing Map}
\acrodef{SPARQL}[SPARQL]{Simple Protocol and RDF Query Language}
\acrodef{SPE}[SPE]{Software Performance Engineering}
\acrodef{SPINA}[SPINA]{Superimposed Peer Infrastructure for iNformation Access}
\acrodef{SPLIT}[SPLIT]{Stemming Program for Language Independent Tasks}
\acrodef{SPOOL}[SPOOL]{Simultaneous Peripheral Operations On Line}
\acrodef{SQL}[SQL]{Structured Query Language}
\acrodef{SR}[SR]{Sliding Ratio}
\acrodef{sRBP}[sRBP]{Session Rank Biased Precision}
\acrodef{SRU}[SRU]{Search/Retrieve via \acs{URL}}
\acrodef{SS}[SS]{Sum of Squares}
\acrodef{SSTF}[SSTF]{Shortest Seek Time First}
\acrodef{STAR}[STAR]{Steiner-Tree Approximation in Relationship graphs}
\acrodef{STON}[STON]{STemming ON}
\acrodef{SVM}[SVM]{Support Vector Machine}
\acrodef{TAC}[TAC]{Text Analysis Conference}
\acrodef{TBG}[TBG]{Time-Biased Gain}
\acrodef{TCP}[TCP]{Transmission Control Protocol}
\acrodef{TEL}[TEL]{The European Library}
\acrodef{TERRIER}[TERRIER]{TERabyte RetrIEveR}
\acrodef{TF}[TF]{Term Frequency}
\acrodef{TFR}[TFR]{True False Rate}
\acrodef{TLD}[TLD]{Top Level Domain}
\acrodef{TME}[TME]{Topics--Metrics-Experiments}
\acrodef{TN}[TN]{True Negative}
\acrodef{TO}[TO]{Transfer Object}
\acrodef{TP}[TP]{True Positve}
\acrodef{TPR}[TPR]{True Positive Rate}
\acrodef{TRAT}[TRAT]{Text Relevance Assessing Task}
\acrodef{TREC}[TREC]{Text REtrieval Conference}
\acrodef{TRECVID}[TRECVID]{TREC Video Retrieval Evaluation}
\acrodef{TTL}[TTL]{Time-To-Live}
\acrodef{UCD}[UCD]{Use Case Diagram}
\acrodef{UDC}[UDC]{Universal Decimal Classification}
\acrodef{uGAP}[uGAP]{User-oriented Graded Average Precision}
\acrodef{UI}[UI]{User Interface}
\acrodef{UML}[UML]{Unified Modeling Language}
\acrodef{UMT}[UMT]{User Management Tool}
\acrodef{UMTS}[UMTS]{Universal Mobile Telecommunication System}
\acrodef{UoM}[UoM]{Utility-oriented Measurement}
\acrodef{UPC}[UPC]{Universal Product Code}
\acrodef{URI}[URI]{Uniform Resource Identifier}
\acrodef{URL}[URL]{Uniform Resource Locator}
\acrodef{URN}[URN]{Uniform Resource Name}
\acrodef{USM}[USM]{User Storing Manager}
\acrodef{VA}[VA]{Visual Analytics}
\acrodef{VAIRE}[VAIR\"{E}]{Visual Analytics for Information Retrieval Evaluation}
\acrodef{VATE}[VATE$^2$]{Visual Analytics Tool for Experimental Evaluation}
\acrodef{VIRTUE}[VIRTUE]{Visual Information Retrieval Tool for Upfront Evaluation}
\acrodef{VD}[VD]{Virtual Document}
\acrodef{VDM}[VDM]{Visual Data Mining}
\acrodef{VIAF}[VIAF]{Virtual International Authority File}
\acrodef{VIM}[VIM]{International Vocabulary of Metrology}
\acrodef{VL}[VL]{Visual Language}
\acrodef{VoIP}[VoIP]{Voice over IP}
\acrodef{VS}[VS]{Visual Sentence}
\acrodef{W3C}[W3C]{World Wide Web Consortium}
\acrodef{WAN}[WAN]{Wide Area Network}
\acrodef{WHO}[WHO]{World Health Organization}
\acrodef{WLAN}[WLAN]{Wireless \acs{LAN}}
\acrodef{WP}[WP]{Work Package}
\acrodef{WS}[WS]{Web Services}
\acrodef{WSD}[WSD]{Word Sense Disambiguation}
\acrodef{WSDL}[WSDL]{Web Services Description Language}
\acrodef{WWW}[WWW]{World Wide Web}
\acrodef{XMI}[XMI]{\acs{XML} Metadata Interchange}
\acrodef{XML}[XML]{eXtensible Markup Language}
\acrodef{XPath}[XPath]{XML Path Language}
\acrodef{XSL}[XSL]{eXtensible Stylesheet Language}
\acrodef{XSL-FO}[XSL-FO]{\acs{XSL} Formatting Objects}
\acrodef{XSLT}[XSLT]{\acs{XSL} Transformations}
\acrodef{YAGO}[YAGO]{Yet Another Great Ontology}
\acrodef{YASS}[YASS]{Yet Another Suffix Stripper}

\clearpage
\bibliographystyle{ACM-Reference-Format}


\balance

\bibliography{main}  



\end{document}